\documentclass[usenatbib,times,usegraphicx,useAMS]{mn2e}
\usepackage{amssymb,url,times}

\newcommand{\pder}[2]{\ensuremath{\frac{\partial #1}{\partial #2}}}

\defcitealias{LP07}{LP07} 
\defcitealias{ogilvie99}{O99} 

\title[Warp diffusion in thin discs]{On the diffusive propagation of warps in thin accretion discs}

\author[Lodato \& Price]{Giuseppe Lodato$^{1,2}$\footnote{giuseppe.lodato@unimi.it}, Daniel J. Price$^{3,4}$ \\
$^1$Dipartimento di Fisica, Universit\`a Degli Studi di Milano, Via Celoria, 16, Milano, 20133, Italy\\
$^2$  Isaac Newton Institute for Mathematical Studies, 20 Clarkson Road, Cambridge CB3 0EH \\
$^3$Centre for Stellar and Planetary Astrophysics, School of Mathematical Sciences, Monash University, Clayton 3800, Australia \\
$^4$School of Physics, University of Exeter, Stocker Rd, Exeter EX4 4QL
}
\date{Submitted: Revised:  Accepted:}
\pagerange{\pageref{firstpage}--\pageref{lastpage}} \pubyear{2010}

\begin{document}
\label{firstpage}
\bibliographystyle{mn2e}
\maketitle

\begin{abstract}

In this paper we revisit the issue of the propagation of warps in thin and viscous accretion discs. In this regime warps are know to propagate diffusively, with a diffusion coefficient approximately inversely proportional to the disc viscosity. Previous numerical investigations of this problem \citep{LP07} did not find a good agreement between the numerical results and the predictions of the analytic theories of warp propagation, both in the linear and in the non-linear case. Here, we take advantage of a new, low-memory and highly efficient Smoothed Particle Hydrodynamics (SPH) code to run a large set of very high resolution simulations (up to 20 million SPH particles) of warp propagation, implementing an isotropic disc viscosity in different ways, to investigate the origin of the discrepancy between the theory and the numerical results. We identify the cause of the discrepancy in an incorrect calibration of disc viscosity in \citet{LP07}. Our new and improved analysis now shows a remarkable agreement with the analytic theory both in the linear and in the non-linear regime, in terms of warp diffusion coefficient and precession rate. It is worth noting that the resulting diffusion coefficient is inversely proportional to the disc viscosity only for small amplitude warps and small values of the disc $\alpha$ coefficient ($\alpha\lesssim 0.1$). For non-linear warps, the diffusion coefficient is a function of both radius and time, and is significantly smaller than the standard value. Warped accretion discs are present in many contexts, from protostellar discs to accretion discs around supermassive black holes. In all such cases, the exact value of the warp diffusion coefficient may strongly affect the evolution of the system and therefore its careful evaluation is critical in order to correctly estimate the system dynamics. 

\end{abstract}

\begin{keywords}
accretion, accretion discs --- hydrodynamics --- instabilities
\end{keywords}

%----------------------------------------------------------------------------------------------------------------
\section{Introduction}

Warped accretion discs may occur across a wide variety of astrophysical systems, from the large scales of accretion discs around supermassive black holes (SMBH), down to the small scales of planet forming discs. 

Observationally, warps are found in galactic binary systems, such as the hyperaccreting X-ray binary SS433 \citep{begelman06b} and the X-ray binary Her X-1 \citep{wijers99}, and in several microquasars, including GRO J1655-40 \citep{martin08a} and V4641 Sgr \citep{martin08b}. On the much less energetic side, a warped protostellar disc is found around the young star KH 15D \citep{chiang04}. 

Warps are also found in the thin accretion discs in Active Galactic Nuclei (AGN), as in the case of NGC 4258 \citep{herrnstein96,papaloizou98}. The dynamics of warped accretion can play a fundamental role in these cases, as it in turn regulates the spin history of the growing SMBH and, as a consequence, its very ability to grow rapidly \citep{king06,KPH08}.

The torques which produce the warp can be very different. For protostellar discs, they include tidal interactions with a companion star \citep{larwood96,martin09}, and dynamical effects during the formation of the disc, which might affect the relative orientation of the stellar spin and the planetary orbits \citep{BLP09}. For accretion discs around black holes there are additional torques arising from the general relativistic Lense-Thirring precession around a spinning black hole \citep{bardeen75,scheuer96,KLOP,LP06,martin07b,perego09}, and self-induced warping caused by radiation pressure \citep{pringle96}. Recently, some attention has also been given to the process of disc warping and black hole spin alignment in the case of supermassive black hole binaries \citep{dotti09}. In all such cases, the evolution of the system is strongly dependent on the speed at which warping disturbances can propagate in the disc.

Analytic theories of warp propagation have been discussed extensively in the past \citep{pappringle83,pringle92,paplin95,ogilvie99,ogilvie00} (see Section \ref{sec:theory}). These theories predict that while for thick discs warps should propagate as dispersive waves, with a velocity of the order of half the sound speed in the disc, in the limit of thin and viscous discs the propagation is diffusive, with a diffusion coefficient inversely proportional to the disc viscosity \citep{pappringle83}. Numerical simulations of warp propagation in the thick disc case have been performed by \citet{nelson99}  and \citet{nelson00}. 

Numerical simulations of warp propagation for thin and viscous discs are much more challenging, because in order to properly catch the warp dynamics it is essential to accurately resolve the vertical structure of the disc, which for very thin discs can be difficult. A first attempt at testing the analytical theory with numerical simulations in the thin disc regime has been performed by \citet{LP07} (hereafter \citetalias{LP07}), using SPH. The results of \citetalias{LP07} showed some unexpected results: while for large values of the disc viscosity the warp diffusion coefficient appeared to scale inversely with viscosity, as predicted analytically, such behaviour was not found at low viscosities, \emph{already for small warp amplitudes}. In this case, the diffusion coefficient appeared to be much smaller than theoretically predicted, implying (as discussed extensively by \citetalias{LP07}) some additional dissipation. Additionally, the internal precession induced by the warp and predicted analytically was found to be strongly dependent on the specific implementation of viscosity and was not found to match the theoretical expectations. 

\citetalias{LP07} discuss different possible explanations for such disagreement. On the one hand, it is quite possible that the limited numerical resolution of their simulations might have affected their results. On the other hand, strong supersonic motions were found in the \citetalias{LP07} simulations, which might result into shocks in the resulting flow and thus provide the required additional dissipation.

In this paper, we want to systematically address all the issues left open by \citetalias{LP07}, by checking both the numerical aspects of the problem and the physical effects involved. 

 With regards to the numerical aspects, first of all we have used a different SPH code with respect to \citetalias{LP07}, therefore validating one code against the other. Secondly, we have checked numerical convergence by running simulations using 20 million particles, that is a factor of ten larger than \citetalias{LP07} (note that such simulations are among the largest SPH simulations of accretion discs performed to date). Since some of the effects reported by \citetalias{LP07} appeared to depend on the viscosity formulation, we have here tested two different possible implementations of disc viscosity. Finally, we have modified our analysis procedure, so as to obtain a more quantitative evaluation of the uncertainties in the measured parameters. In order to test the physical effects which might determine the \citetalias{LP07} results, we have paid attention to shocks, which were argued by \citetalias{LP07} to be responsible for the additional dissipation. We have checked the importance of shocks by running simulations with different levels of bulk viscosity --- varied independently of the shear viscosity --- which is directly connected with shock dissipation.

The paper is organised as follows. In section \ref{sec:theory} we discuss the basic features of the analytic theory of warp propagation in both the linear and non-linear regime. In section \ref{sec:numerics} we detail the numerical method that we have used to simulate the system, including the different implementations of disc viscosity that we use. In section \ref{sec:analysis} we describe the procedure we have used to analyse our results and extract from the simulations the warp diffusion parameters. In section \ref{sec:results} we present and discuss our main results for the warp diffusion and precession in both the linear and non-linear regime. Finally, in section \ref{sec:conclusions} we draw our conclusions.

\section{Analytic theories of warp propagation}
\label{sec:theory}
We consider here, as in \citetalias{LP07}, the propagation of warps in thin Keplerian accretion discs,
rotating with angular velocity $\Omega(R)$, with surface density $\Sigma(R)$
and angular momentum per unit area ${\bf L}(R)$. Here $R$ should be
interpreted as a `spherical' coordinate. The local direction of ${\bf L}$ can
be oriented arbitrarily in space, and the unit vector ${\bf l}(R)={\bf L}(R)/L(R)$
defines its direction. If the disc is rotating around a central point mass
$M$, then its rotation is Keplerian, with $\Omega=\sqrt{GM/R^3}$ and
$L(R)=\Sigma(R)\sqrt{GMR}$.

The disc is warped whenever the direction identified by ${\bf l}$ changes with
radius. The warp amplitude may be characterised using the dimensionless
parameter $\psi$, where
\begin{equation}
\psi=R\left|\frac{\partial{\bf l}(R)}{\partial R}\right|.
\end{equation}

The disc thickness is $H=c_{\rm s}/\Omega$, where $c_{\rm s}$ is the sound
speed, and is the scale over which density and pressure change in the local
$z$ direction. The disc aspect ratio is $H/R$, and we shall assume that
$H/R\ll 1$.

We use the standard \citet{shakura73} prescription for the disc viscosity $\nu$, assumed here to be a standard, isotropic, Navier-Stokes viscosity:
\begin{equation}
\nu=\alpha c_{\rm s}H = \alpha\Omega H^2.
\label{eq:shakura}
\end{equation}

Warping disturbances can propagate in accretion discs in two different regimes, depending on the relative importance of pressure forces and viscous forces. If the disc is sufficiently thick, such that $H/R>\alpha$, then the warp propagates as a dispersive wave \citep{paplin95}. The equations of motion for a wave in the case where the disc is Keplerian and
nearly inviscid are \citep{lubow00,lubow02}

\begin{equation}
\Sigma R^3\Omega \frac{\partial{\bf l}}{\partial t} = \frac{\partial{\bf G}}
{\partial R}
\label{eq:wave1}
\end{equation}
\begin{equation}
\frac{\partial{\bf G}}{\partial t} + \alpha\Omega{\bf G} =
\Sigma R^3\Omega \frac{c_{\rm s}^2}{4}\frac{\partial{\bf l}}{\partial R},
\label{eq:wave2}
\end{equation}
where ${\bf G}$ is the disc internal torque in the horizontal plane (only). Note that these equations are valid only in the linear approximation for small warps, and that no general non-linear theory for the wave-like regime exists as yet (but see \citealt{ogilvie06}).

Here, we are mostly interested in the case where the disc is thin and viscous, such that $H/R <«\alpha$. In this case, the warp propagates diffusively \citep{pappringle83}, and can be approximately described by the equation
\citep{pringle92}
\begin{eqnarray}
\label{eq:pringle}
\nonumber\frac{\partial{\bf L}}{\partial
t} = &&\frac{3}{R}\frac{\partial}{\partial R}\left[\frac{R^{1/2}}{\Sigma}
\frac{\partial}{\partial R}
(\nu_1\Sigma R^{1/2}){\bf L}\right]\\
&+&\frac{1}{R}\frac{\partial}{\partial R}\left[\left(\nu_2 R^2\left|
\frac{\partial{\bf l}}{\partial R}\right|^2-\frac{3}{2}\nu_1\right)
{\bf L}\right]\\
\nonumber&+&\frac{1}{R}\frac{\partial}{\partial
R}\left(\frac{1}{2}\nu_2R|{\bf L}|\frac{\partial{\bf l}}{\partial R}
\right).
\end{eqnarray}
In this equation, the terms proportional to $\nu_1$ describe the standard
viscous evolution of a thin and flat disc. For small amplitude warps, $\nu_1=\nu$ (and thus $\alpha_1=\alpha$), but for large amplitudes $\nu_1$ can be affected by the warp. The terms proportional to $\nu_2$ arise whenever the disc is warped and $|\partial{\bf l}/\partial R|\neq 0$. According to Equation (\ref{eq:pringle}) the warp diffuses with a diffusion
coefficient $\nu_2$. By analogy with the viscosity prescription (Eq. (\ref{eq:shakura})) we
can define a second parameter $\alpha_2$ so that
\begin{equation}
\nu_2=\alpha_2 c_{\rm s}H = \alpha_2\Omega H^2.
\end{equation}

It is clear that the nature of the evolution of a warped accretion disc is
determined mainly by the relative values of $\alpha$ and $\alpha_2$.
In the case of small warp amplitude, $\psi \ll H/R$, and for viscosity such
that $H/R \lesssim \alpha \ll 1$, \citet{pappringle83} have found the following
relation between the two coefficient $\alpha$ and $\alpha_2$:
\begin{equation}
\label{eq:prediction}
\alpha_2 = \frac{1}{2 \alpha},
\end{equation}
and therefore that the warp diffusion coefficient is {\it inversely}
proportional to the size of the viscosity. \citet{ogilvie99} (hereafter \citetalias{ogilvie99})  extends these
approximate analytic results by use of an asymptotic expansion in terms of the
small quantity $H/R$, but retaining the assumption of an isotropic
(Navier-Stokes) viscosity. By this means he is able to take account of larger
values of $\alpha$ and $\psi$. In the limit of a small amplitude warp ($\psi \ll 1$),  
\citetalias{ogilvie99} finds
\begin{equation}
\frac{\nu_2}{\nu_1}=\frac{1}{2\alpha^2}\frac{4(1+7\alpha^2)}{4+\alpha^2},
\label{eq:prediction2}
\end{equation}
which includes higher order corrections in $\alpha$. \citetalias{ogilvie99} also computes the relation between $\alpha$ and $\alpha_2$ for an arbitrarily large warp amplitude, for which there is no simple analytical expression, but that can be computed numerically. In order to analyse our results we will also make use of these numerical relations (see Sec. \ref{sec:results}).

Finally, it should be noted that the full non-linear theory of \citetalias{ogilvie99} also includes some precessional torques, which are not  accounted for in the diffusion model by \citet{pringle92}, because they arise at higher order in $\alpha$. In interpreting our results, in some cases, we have added such terms in our simple diffusion model, by adding a term on the right-hand side of equation
(\ref{eq:pringle}), in the form of (see \citealt{ogilvie99})
\begin{equation}
\left.\frac{\partial{\bf L}}{\partial t}\right |_{\rm prec} =  
\frac{1}{R}\frac{\partial}{\partial
R}\left(\nu_3R|{\bf L}| {\bf l}\times\frac{\partial{\bf l}}{\partial R}
\right),
\label{eq:precession}
\end{equation}
where we have introduced a third coefficient $\nu_3$ related to precessional effects (with a corresponding $\alpha_3=\nu_3/\Omega H^2$). The non-linear theory of \citetalias{ogilvie99} also provides an expression for the dependence of $\alpha_3$ on $\alpha$ and on $\psi$. In the limit of a small amplitude warp ($\psi \ll 1$) and to lowest non-zero order in $\alpha$, the precession coefficient is given by \citepalias{ogilvie99}\footnote{Note that this is given incorrectly as 3/4 in \citet{pappringle83} and correspondingly in \citetalias{LP07}.}
\begin{equation}
\alpha_{3} = 3/8.
\label{eq:alpha3const}
\end{equation}
Taking account of large values of $\alpha$, but still for small amplitude warps, the expected relation is given by (cf. \citealt{ogilviedubus}, Eq. 12)
\begin{equation}
\alpha_{3} = \frac{3(1 - 2\alpha^{2})}{2(4 + \alpha^{2})}.
\label{eq:alpha3vsalpha}
\end{equation}
For non-linear warp amplitudes, higher order corrections in both $\alpha$ and $\psi$ are given by \citetalias{ogilvie99}.\footnote{The reader should note that \citetalias{ogilvie99} uses a slightly different notation, defining the coefficients using $Q_1 \equiv -3\alpha_1/2$, $Q_2 \equiv \alpha_2/2$ and $Q_3 \equiv \alpha_3$.}

\section{Numerical method}
\label{sec:numerics}
 We have performed a series of three-dimensional SPH simulations of warped discs similar, though more extensive than, those performed by \citetalias{LP07}. SPH is a Lagrangian scheme for solving the equations of hydrodynamics in which fluid quantities and their derivatives are computed on a set of $N$ particles that follow the fluid motion (see \citealt{price04} or \citealt{monaghan05} for recent reviews).

 In this paper we have used the \textsc{phantom} code, developed by D. Price (see \citealt{pf10} for another recent application). \textsc{phantom} is a low-memory, highly efficient SPH code optimised for studying non-self-gravitating problems. The code is made very efficient by using a simple neighbour finding scheme based on a fixed (in this case, cylindrical) grid and linked lists of particles. In particular, the absence of overheads associated with the tree-code for computing gravitational forces as well as other optimisations means that the code is significantly more efficient than the \citet{benzetal90} / \citet{batephd}-derived code previously employed by \citetalias{LP07}.

 The initial aim of using \textsc{phantom} was, given the concerns in \citetalias{LP07}, to be able to employ a much higher resolution in the calculations. Whilst in the end we found this unnecessary (see Sec. \ref{sec:results}), we have instead used our increased computing ability to survey a much wider parameter space than that explored by \citetalias{LP07}, including a wide range of viscosity parameters, two different viscosity formulations and three different warp amplitudes.

\subsection{Navier-Stokes equations}
\label{sec:ns}
 In this paper we compute the evolution of a viscous accretion disc by solving the Navier-Stokes equations for a viscous, compressible hydrodynamic gas in an external potential, given by
\begin{eqnarray}
\frac{d\rho}{dt} & = & -\rho \nabla\cdot{\bf v}, \label{eq:cty} \\
\frac{dv^{i}}{dt} & =& -\frac{1}{\rho}\pder{S^{ij}}{x^{j}} + f^{i}_{\rm pot}, \label{eq:mom}
\end{eqnarray}
where the potential corresponds to the gravitational force from a central star or black hole of mass $M$ at the origin,  i.e.,
\begin{equation}
{\bf f}_{\rm pot} = -\frac{GM}{r^{2}} \hat{\bf r},
\end{equation}
and the stress tensor is given by the usual expression
\begin{equation}
S^{ij} = \left[-P + \left(\zeta - \frac23 \eta\right)\pder{v^{k}}{x^{k}}\right] \delta^{ij} + \eta \left( \pder{v^{i}}{x^{j}} + \pder{v^{j}}{x^{i}}\right),
\label{eq:sij}
\end{equation}
where $\eta$ and $\zeta$ are the shear and bulk viscosity coefficients respectively. Note that the kinematic shear viscosity $\nu$ is related to the shear viscosity $\eta$ by $\nu = \eta/\rho$ and similarly one may define the volume viscosity $\zeta_{v} = \zeta/\rho$. In the case of constant viscosity coefficients, the equations can be simplified to the vector form
\begin{equation}
\frac{d{\bf v}}{dt} = -\frac{\nabla P}{\rho} + \frac{\eta \nabla^{2} {\bf v}}{\rho} + \left( \zeta + \frac{\eta}{3} \right) \frac{\nabla (\nabla\cdot{\bf v})
}{\rho} + {\bf f}_{\rm pot}.
\label{eq:nsvec}
\end{equation}

The pressure is related to the density via a locally isothermal equation of state
\begin{equation}
P = c_{\rm s}^{2}(R) \rho,
\end{equation}
where the sound speed $c_{s}$ is a prescribed function of (spherical) radius $R \equiv \sqrt{x^{2} + y^{2} + z^{2}}$, given by
\begin{equation}
c_{\rm s}(R) = c_{\rm s,0} R^{-q},
\label{eq:cs}
\end{equation}
where $q $ is given in Section \ref{sec:ics} and the normalisation $c_{s,0}$ determines the disc thickness.

 In the \citet{shakura73} $\alpha$-parametrisation for the disc viscosity, the kinematic viscosity coefficient is given by Eq. (\ref{eq:shakura}). We consider two different methods for implementing Navier-Stokes viscosity in SPH, firstly based on a modification to the usual artificial viscosity term (similar to that employed by \citetalias{LP07}) and secondly based on a direct evaluation of the derivatives in (\ref{eq:sij}) and (\ref{eq:mom}) respectively (see Sec. \ref{sec:flebbe}). The procedure used in the former case is described in Sec. \ref{sec:avvisc}, setting $\eta$ and $\zeta$ to zero in Eq. \ref{eq:sij}. For the latter case we simply specify $\nu$ in (\ref{eq:sij}) from the nominally input value for $\alpha$ using
\begin{equation}
\nu(R) = \alpha\frac{c_{\rm s}^{2}(R)}{\Omega(R)},
\label{eq:nualpha}
\end{equation}
where $c_{s}(R)$ is specified according to Eq. (\ref{eq:cs}) and we assume a Keplerian rotation profile $\Omega = \sqrt{GM/R^{3}}$. The corrections to Keplerian rotation due to the pressure gradient are of order $(H/R)^{2}$, which for the thin discs considered in this paper are very small ($\sim 10^{-4}$).

\subsection{SPH}
\label{sec:sph}
\subsubsection{Hydrodynamics}
\label{sec:sphhydro}
 \textsc{phantom} implements the full variable smoothing length SPH formulation developed by \citet{pm04b} and \citet{pm07}, whereby the smoothing length, $h$, and density, $\rho$, are mutually dependent via the density sum (for particle $a$)
\begin{equation}
\rho_{a} = \sum_{b} m_{b} W_{ab} (h_{a}),
\label{eq:rhosum}
\end{equation}
which is an exact solution to (\ref{eq:cty}), and the relation
\begin{equation}
h_{a} = h_{\rm fac} \left(\frac{m_{a}}{\rho_{a}}\right)^{1/3},
\label{eq:hrho}
\end{equation}
where $m_a$ is the particle mass and $W_{ab} \equiv W(\vert {\bf r}_{a} -  {\bf r}_{b}\vert, h_{a})$ is the SPH smoothing kernel (see e.g. \citealt{monaghan92,price04,monaghan05} for details). This results in a resolution that adapts to the local particle number density. Equations (\ref{eq:rhosum}) and (\ref{eq:hrho}) are iterated self-consistently using a Newton-Raphson method as described in \citet{pm07}, where in this paper we have used $h_{\rm fac}=1.2$, giving approximately 58 neighbours per particle in a smooth distribution.

 The equations of motion (\ref{eq:mom}) take the form
\begin{eqnarray}
\frac{dv^{i}_{a}}{dt} &= & -\sum_{b} m_{b} \left[ \frac{S^{ij}_{a}}{\Omega_{a}\rho_{a}^{2}} \nabla_{a}^{j} W_{ab} (h_{a}) + \frac{S^{ij}_{b}}{\Omega_{b}\rho_{b}^{2}} \nabla_{a}^{j} W_{ab} (h_{b})\right], \nonumber \\
& & + f^{i}_{\rm pot},
\label{eq:sphmom}
\end{eqnarray}
where $\Omega$ is a dimensionless quantity related to the smoothing length gradients (see \citealt{pm07} for details) and the stress tensor is given by
\begin{eqnarray}
S^{ij}_{a} &=& \left[-\left(P_{a} + q^{AV}_{a}\right) + \left(\zeta_{a} - \frac23 \eta_{a}\right)\pder{v^{k}_{a}}{x^{k}_{a}}\right] \delta^{ij} \nonumber \\
&+& \eta_{a} \left( \pder{v^{i}_{a}}{x^{j}_{a}} + \pder{v^{j}_{a}}{x^{i}_{a}}\right). \label{eq:sijsph}
\end{eqnarray}

The $q^{AV}$ term in (\ref{eq:sijsph}) is the artificial viscosity (discussed below) which is introduced in SPH in order to capture shocks and (to a lesser extent) to prevent interpenetration of particles. However, it can be shown (see below) that the artificial viscosity corresponds directly to a Navier-Stokes type term and can thus be used, with minor adjustment of the parameters, to directly represent physical viscous diffusion (in doing so one would obviously discard the remaining terms in Eq. \ref{eq:sijsph}, i.e., setting $\zeta=\eta=0$). The disadvantage of doing so is that the resultant viscosity coefficient consists of both shear and bulk components of viscosity, whereas for a disc the viscosity parameterisation (\ref{eq:shakura}) should consist of shear viscosity only.

The remaining part of the SPH algorithm is the time integration algorithm, for which we use a standard leapfrog scheme equivalent to the velocity Verlet method. For efficiency we assign individual timesteps, set in factors of 2 from a nominal maximum timestep, such that only a subset of the particles is moved on the shortest timestep. With individual timesteps many of the conservation properties of the leapfrog algorithm are only approximately satisfied, however the scheme is significantly more efficient.

\subsubsection{Artificial viscosity}
\label{sec:av}
 The artificial viscosity formulation in \textsc{phantom} follows that of \citet{monaghan97}, with the averaging in the density and signal velocity changed slightly in order to more efficiently calculate the terms in (\ref{eq:sphmom}). We use
\begin{equation}
q^{AV}_{a} = \left\{
\begin{array}{ll}
\frac12 \alpha^{\rm AV}_{a} \rho_{a} v_{{\rm sig},a} \vert {\bf v}_{ab}\cdot\hat{\bf r}_{ab} \vert,& {\bf v}_{ab}\cdot\hat{\bf r}_{ab} < 0 \\
0 &  {\bf v}_{ab}\cdot\hat{\bf r}_{ab} \ge 0
\end{array}\right.
\label{eq:qvisc}
\end{equation}
where ${\bf v}_{ab} \equiv {\bf v}_{a} - {\bf v}_{b}$ and the viscosity is only applied for approaching particles (${\bf v}_{ab}\cdot\hat{\bf r}_{ab} < 0$, i.e., converging flows). The signal velocity for hydrodynamics is given by
\begin{equation}
v_{{\rm sig},a} = c_{{\rm s},a} + \beta^{\rm AV} \vert {\bf v}_{ab}\cdot\hat{\bf r}_{ab} \vert,
\label{eq:vsig}
\end{equation}
where in general $\beta^{\rm AV} = 2$. The $\beta^{\rm AV}$ term in the signal velocity provides a non-linear term that was originally introduced to prevent particle penetration in high Mach number shocks (see e.g. \citealt{monaghan89}).

 For shock capturing --- where the aim is to provide as little dissipation as possible whilst resolving shock structures --- \textsc{phantom} implements the \citet{mm97} switch to reduce dissipation away from shocks, in which the dissipation parameter $\alpha^{\rm AV}$ is evolved according to a source and decay equation
\begin{equation}
\frac{d\alpha^{\rm AV}_{a}}{dt} = - \frac{\alpha^{\rm AV}_{a} - \alpha^{\rm AV}_{\rm min}}{\tau_{a}} + \mathcal{S}_{a}, \hspace{1cm} \tau_{a} = h_{a}/(\sigma c_{\rm s})
\end{equation}
where $\sigma=0.1$, $\mathcal{S} = \max (0,-\nabla\cdot{\bf v})$, $\alpha_{min} = 0.05$ and in general one would enforce $\alpha_{max}=1.0$.

\subsubsection{Disc viscosity using the artificial viscosity term}
\label{sec:avvisc}
 It has been known for quite some time \citep[e.g.][]{lubow94,murray96} that the artificial viscosity terms in SPH can be understood straightforwardly as numerical representations of second derivatives of the velocity. This is because the standard procedure for evaluating second derivatives in SPH is to use an integral formulation based on only the first derivative of the SPH kernel. For example the Laplacian for a scalar quantity $A$ is represented by \citep{brookshaw85,price04,monaghan05}
\begin{equation}
\nabla^{2} A_{a} =  - 2 \sum_{b} \frac{m_{b}}{\rho_{b}} ( A_{a} - A_{b})  \frac{ \hat{\bf r}_{ab}\cdot\nabla_{a} W_{ab}}{\vert r_{ab} \vert},
\end{equation}
which is more clearly expressed by writing the kernel gradient as $\nabla_{a} W_{ab} \equiv \hat{\bf r}_{ab} F_{ab}$, giving
\begin{equation}
\nabla^{2} A =  - 2 \sum_{b} \frac{m_{b}}{\rho_{b}} ( A_{a} - A_{b}) \frac{F_{ab}}{\vert r_{ab} \vert}.
\end{equation}

For a vector quantity the corresponding expressions are \citep{er03,monaghan05}
\begin{eqnarray}
\nabla^{2} {\bf A} & = & - 2 \sum_{b} \frac{m_{b}}{\rho_{b}} ({\bf A}_{a} - {\bf A}_{b}) \frac{F_{ab}}{\vert r_{ab} \vert}, \label{eq:del2A} \\
\nabla (\nabla\cdot {\bf A}) & = & - \sum_{b} \frac{m_{b}}{\rho_{b}} \left[(\delta^{k}_{k} + 2)({\bf A}_{ab}\cdot\hat{\bf r}_{ab})\hat{\bf r
}_{ab} - {\bf A}_{ab} \right]  \frac{F_{ab}}{\vert r_{ab} \vert}, \nonumber \\ \label{eq:graddivA}
\end{eqnarray}
where $\delta^{k}_{k} \equiv n$ i.e., the number of spatial dimensions.

 The above expressions mean that it is possible to give clear interpretation to the artificial viscosity terms, since for example in three dimensions we have, from (\ref{eq:del2A}) and (\ref{eq:graddivA}),
\begin{equation}
-\sum_{b} \frac{m_{b}}{\rho_{b}} ({\bf A}_{ab}\cdot \hat{\bf r}_{ab}) \frac{F_{ab}}{\vert r_{ab} \vert} = \frac{1}{5}\nabla (\nabla\cdot {\bf A}) +\frac{1}{10} \nabla^{2} {\bf A}.
\end{equation}
%%%%%%%
%%% DAN, DO WE NEED THIS PARAGRAPH???
%% GIUSEPPE, YES WE DO, BECAUSE THIS SHOWS THE TERMS ARE CORRECT EVEN WITH NON-CONSTANT COEFFICIENTS
For non-constant coefficients the expressions are similar \citep{cm99,monaghan05}, but with an average of the coefficients on the particles, for example
\begin{equation}
\nabla\cdot (\kappa \nabla {\bf A}) = - \sum_{b} \frac{m_{b}}{\rho_{b}}(\kappa_{a} + \kappa_{b}) ({\bf A}_{a} - {\bf A}_{b}) \frac{F_{ab}}{\vert r_{ab} \vert}, \label{eq:del2Akappa}
\end{equation}
where \citet{cm99} give alternative averaging procedures more appropriate when the coefficients are discontinuous.
%%%%%%%%

\begin{figure}
\begin{center}
\includegraphics[width=\columnwidth]{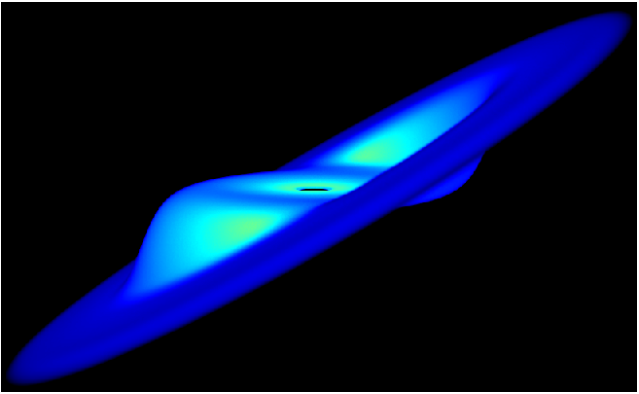}
\caption{3D structure of the warped accretion disc from a representative 20 million particle calculation in the large warp amplitude case ($A=0.5$).}
\label{fig:warp3D}
\end{center}
\end{figure}

Thus, for the artificial viscosity terms presented above (Sec. \ref{sec:av}), we have
\begin{eqnarray}
\nu^{\rm AV} & \approx & \frac{1}{10} \alpha^{\rm AV} v_{\rm sig} \vert r_{ab} \vert, \\
\zeta_{v}^{\rm AV} & \approx & \frac{1}{6} \alpha^{\rm AV} v_{\rm sig} \vert r_{ab} \vert.
\end{eqnarray}
Note that the factor $\vert r_{ab} \vert$ in the \citet{monaghan97} formulation of viscosity used in \textsc{phantom} differs slightly from the factor $h$ that would result from using the older \citet{monaghan92} formulation. The difference is only slight because by definition within the kernel radius $\vert r_{ab} \vert / h \le 2$, but use of the \citet{monaghan97} version avoids the need to account for divergences in the denominator when $\vert r_{ab} \vert \to 0$.

 In order to use the artificial viscosity to represent a \citet{shakura73} disc viscosity, we therefore require several minor changes from the formulation appropriate for shocks given in Sec. \ref{sec:av}. These changes are:
\begin{enumerate}
\item Viscosity should be applied for both approaching and receding particles,
\item The $\beta^{\rm AV}$ term in the signal velocity should be dropped such that $v_{\rm sig} = c_{\rm s}$,
\item $q^{AV}$ should be multiplied by a factor $h/\vert r_{ab} \vert$, similar to the \citet{monaghan92} artificial viscosity scheme, and
\item The \citet{mm97} switch should not be used i.e., $\alpha^{\rm AV}$ should be treated as a constant.
\end{enumerate}
 With these conditions the resultant `artificial viscosity for a disc' is given by
\begin{equation}
q^{AV}_{a} = \frac12 \alpha^{\rm AV} \rho_{a} c_{{\rm s},a} \vert {\bf v}_{ab}\cdot\hat{\bf r}_{ab}\vert \frac{h}{\vert r_{ab}\vert},
\end{equation}
giving
\begin{eqnarray}
\nu^{\rm AV} & \approx & \frac{1}{10} \alpha^{\rm AV} c_{\rm s} h, \\
\zeta_{v}^{\rm AV} & \approx & \frac{1}{6} \alpha^{\rm AV} c_{\rm s} h. \label{eq:bulkav}
\end{eqnarray}
 This is essentially the approach adopted by \citetalias{LP07} and several earlier SPH accretion disc calculations \citep[e.g.][]{lubow94,murray96}, giving, from Eq. (\ref{eq:shakura}),
\begin{equation}
\alpha \approx \frac{1}{10} \alpha^{\rm AV} \frac{\langle h \rangle}{H},
\label{eq:nusph}
\end{equation}
where $\langle h \rangle$ is the azimuthally averaged (or, for a warped disc, shell averaged) smoothing length. The additional complication when using a spatially variable smoothing length, addressed by \citetalias{LP07}, is that in order to obtain a disc evolution corresponding to a single, uniform value of $\alpha$ thoughout the disc, it is necessary to setup the disc with a surface density profile such that $\langle h \rangle/H \approx {\rm const}$. This is discussed in Sec. \ref{sec:ics}, below.
 
 The disadvantage of using the artificial viscosity term to represent physical viscosity is that one inevitably ends up with a large and unwanted coefficient of bulk viscosity (Eq. \ref{eq:bulkav}). For a disc simulation this is not so disadvantageous since in general $\nabla\cdot{\bf v}$ is not large, so although the coefficient is large, the term to which it is applied is small. However given that at least some of the deviations from the analytic theory found in \citetalias{LP07} could possibly be explained by excess dissipation, it is desirable to perform simulations that either have no explicit bulk component or where the bulk viscosity is carefully controlled as would be the case when applied to shocks in the case of the usual artificial viscosity (Sec. \ref{sec:avvisc}).

\subsubsection{Navier-Stokes viscosity implemented via two first derivatives}
\label{sec:flebbe}
 A straightforward alternative to using the artificial viscosity term to model physical viscosity is simply to evaluate the terms in Eq. (\ref{eq:sijsph}) directly. This is essentially the method proposed by \citet{flebbeetal94}. In this paper we evaluate the gradient terms using the standard variable-smoothing-length gradient operator, given by
\begin{equation}
\pder{v^{i}_{a}}{x^{j}_{a}} = \frac{1}{\rho_{a}\Omega_{a}}\sum_{b} m_{b} (v^{i}_{b} - v^{i}_{a}) \pder{W_{ab}(h_{a})}{x^{j}_{a}},
\label{eq:dvterm}
\end{equation}
where the coefficients of viscosity are set as discussed in Sec. \ref{sec:ns}. Using this method one can in principle use zero bulk viscosity, by setting the bulk coefficient to zero and turning off any artificial viscosity terms. The danger with doing so is that any shocks that \emph{are} present will not be treated appropriately and also that there is nothing to prevent particle interpenetration in the SPH scheme. Thus any such calculations should be treated with the appropriate degree of caution. A better approach, and our default when using this formulation, is to set the physical bulk coefficient to zero in (\ref{eq:sijsph}), but to apply a small and carefully controlled amount of \emph{artificial} viscosity to correctly dissipate shocks and approaching particles using the switches described in Sec. \ref{sec:av}. The resulting coefficient of bulk viscosity in this case is, however, much lower than would be applied when using the artificial viscosity to mimic a disc viscosity (see Sec. \ref{sec:avvisc}, above).

\begin{figure*}
\begin{center}
\includegraphics[width=\textwidth]{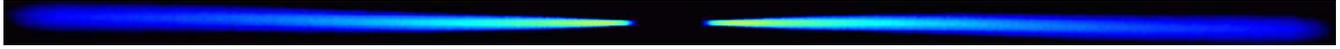}
\caption{Cross section of the disc in the SPH calculations for a low amplitude warp ($A=0.01$) at a resolution of 2 million particles.}
\label{fig:xsection}
\end{center}
\end{figure*}

\subsubsection{Navier-Stokes viscosity using direct second derivatives}
 A further alternative, not considered in this paper, would be to directly evaluate the second derivative terms resulting from the gradient of the stress tensor as in Eq. (\ref{eq:nsvec}), using the standard representation of second derivatives in SPH given by Eqs. (\ref{eq:del2A}) and (\ref{eq:graddivA}). Indeed this forms the basis of the `dissipative particle dynamics' scheme of \citet{er03}. The terms in this case are obviously similar to the formulation using artificial viscosity discussed above, except that the shear and bulk components can be set separately. The disadvantage is that the total angular momentum is no longer conserved because the dissipation is not applied along the line of sight joining the particles (meaning that $\sum_{b} m_{b} {\bf r}_{a} \times d{\bf v}_{a}/dt \neq 0$). How serious a limitation this presents in practice for disc simulations has not been clarified, though it would be worthy of further investigation.

\subsubsection{Azimuthal averaging of SPH results}
In order to compare the results of the 3D SPH simulations with the simple diffusion equation (\ref{eq:pringle}) it is necessary to compute azimuthally-averaged disc quantities from the SPH simulations. We perform this averaging by dividing the simulation domain uniformly in $R$ from $R_{\rm in}$ to $R_{\rm out}$ in $N=350$ spherical shells, with radial width $\Delta = (R_{\rm out}-R_{\rm in})/N$. The disc surface density in shell $i$ is then given by the total mass divided by the disc surface corresponding to each shell, i.e.,
\begin{equation}
\Sigma_i = \frac{\sum_j m_j}{\pi[(R_i+\Delta/2)^2-(R_i-\Delta/2)^2]},
\end{equation}
where the sum is performed over all particles in the shell. The average angular momentum is computed as:
\begin{equation}
{\bf J}_i = \frac{\sum_j m_j {\bf r}_j\times {\bf v}_j} {N_i},
\end{equation}
where $N_i$ is the number of particles in shell $i$. Finally, the local direction of the angular momentum vector can be computed using
\begin{equation}
{\bf l}_i = \frac{{\bf J}_i}{|{\bf J_i|}} .
\end{equation}
Examples of the resulting one-dimensional disc profiles are shown in Figures~\ref{fig:fitalpha}, \ref{fig:wave}, \ref{fig:profilelargeA}, \ref{fig:profilelargeAsmallalpha}, \ref{fig:precession} and \ref{fig:prec_largeA}. The above procedure for computing the disc surface density has also been implemented as a feature in \textsc{splash} \citep{splashpaper}.

\subsection{Initial conditions}
\label{sec:ics}
Initial conditions are identical to those in \citetalias{LP07}. We use code units in which the gravitational constant $G=1$, the central point mass $M=1$ and the time unit is such that at a radius $R=1$ (in code units) the dynamical time is $\Omega^{-1}=1$.  We place the gas particles in Keplerian orbits in the gravitational potential of a point mass with $M=1$ (in code units). The gas particles are removed from the calculation inside a radius $R=0.5$ (in code units). We distribute the particles using a Monte Carlo placement method such that the disc has a prescribed initial surface density
profile, as described below. The particles are distributed in $z$ so as to attain a Gaussian density profile in the vertical direction, with thickness $H=c_{\rm s}/\Omega$. The random particle placement, whilst simple, means that some settling of the disc occurs during the first few dynamical times of the simulation.

 The orbit of each particle is tilted such that the components of the unit vector ${\bf l}$ are given by
\begin{equation}
l_{x}= \left\{ \begin{array}{ll}
    \displaystyle 0  & \hspace{0.1cm}
\mbox{for} \hspace{0.1cm} R<R_1 \\
\\
    \displaystyle
    \frac{A}{2}\left[1+\sin\left(\pi\frac{R-R_0}{R_2-R_1}\right)\right]
      & \hspace{0.1cm}
\mbox{for} \hspace{0.1cm} R_1<R<R_2\\
\\
    \displaystyle A  & \hspace{0.1cm}
\mbox{for} \hspace{0.1cm} R>R_2 
\end{array}\right.
\label{eq:setup}
\end{equation}
\begin{equation}
l_{\bf y} = 0,
\label{eq:initial}
\end{equation}
\begin{equation}
l_{z} = \sqrt{1-l_{x}^2},
\end{equation}
where $R_1=3.5$ and $R_2=6.5$ in code units, and $R_0 = (R_1 +R_2)/2 = 5$. The warp
amplitude $\psi$ is then
\begin{equation}
  \psi=R\left|\frac{\partial{\bf l}}{\partial R}\right|=\frac{R}{l_{z}}
\frac{\partial l_{x}}{\partial R}
\label{eq:psi}
\end{equation}
the maximum of which is attained at $R\approx R_0$ and is given by
\begin{equation}
\psi_{\rm max}\approx \frac{\pi R_0}{2(R_1-R_2)}\frac{A}{\sqrt{1-(A/2)^2}} = 2.62 \frac{A}{\sqrt{1-(A/2)^2}}
\label{eq:psimax}
\end{equation}

A three dimensional rendering of the resulting warped disc from one of our high resolution (20 million particle) calculations is shown in Figure~\ref{fig:warp3D}, for the case of a high amplitude warp ($A=0.5$). A cross section of the disc in a 2 million particle simulation with a low amplitude warp ($A=0.01$) is shown in Figure~\ref{fig:xsection}, showing the slight bend induced in the disc profile. The initial shape of the warp is also plotted in Fig. 1 of \citetalias{LP07} and is shown by the lines corresponding to $t=0$ in Figures~\ref{fig:wave} and \ref{fig:profilelargeA} of this paper.

The disc extends from $R_{\rm in}=0.5$ to $R_{\rm out}=10$, with a surface
density profile, $\Sigma$, given by
\begin{equation}
\Sigma(R)=\Sigma_0 R^{-p}\left(1-\sqrt{\frac{R_0}{R}}\right).
\end{equation}
The parameter $p$ is set to $p=3/2$, as in \citetalias{LP07}, such that, giving $q=3/4$ in Eq. (\ref{eq:cs}), the disc is uniformly resolved, in the sense that the smoothing length $h\propto \rho^{-1/3}$ is proportional to the disc thickness $H\propto R^{3/4}$, as described in \citetalias{LP07}. 

We choose the normalisation of the sound speed such that the aspect ratio of the disc at $R_0=5$ is $H/R= 0.0133$, corresponding to an aspect ratio at $R=1$ (in code units) of $H/R=0.02$, in order to model a thin disc, where warps propagate primarily in the diffusive regime (see Section \ref{sec:theory}).

\section{Analysis}
\label{sec:analysis}
\subsection{1D Disc evolution}
\label{sec:discevol}

We compare the time evolution of $\Sigma$ and ${\bf l}$ from the SPH simulation with the one resulting from Equation (\ref{eq:pringle}). This is solved using standard finite difference methods, which are detailed in \citet{pringle92} and \citetalias{LP07}, but with a different implementation of the zero torque boundary condition at the inner edge. While in \citet{pringle92} and \citetalias{LP07} the zero torque condition is implemented in an approximate way, by artificially removing mass from the innermost cells in order to keep $\Sigma$ close to zero, in this paper we directly enforce $\Sigma=0$ at the innermost cell. The two conditions are largely equivalent, but the shape of the surface density profile in the inner disc can be slightly modified. This is important because in turn it significantly affects the evaluation of $\alpha$, as described in Section \ref{sec:calib}.

\begin{figure*}
\begin{center}
\includegraphics[width=\columnwidth]{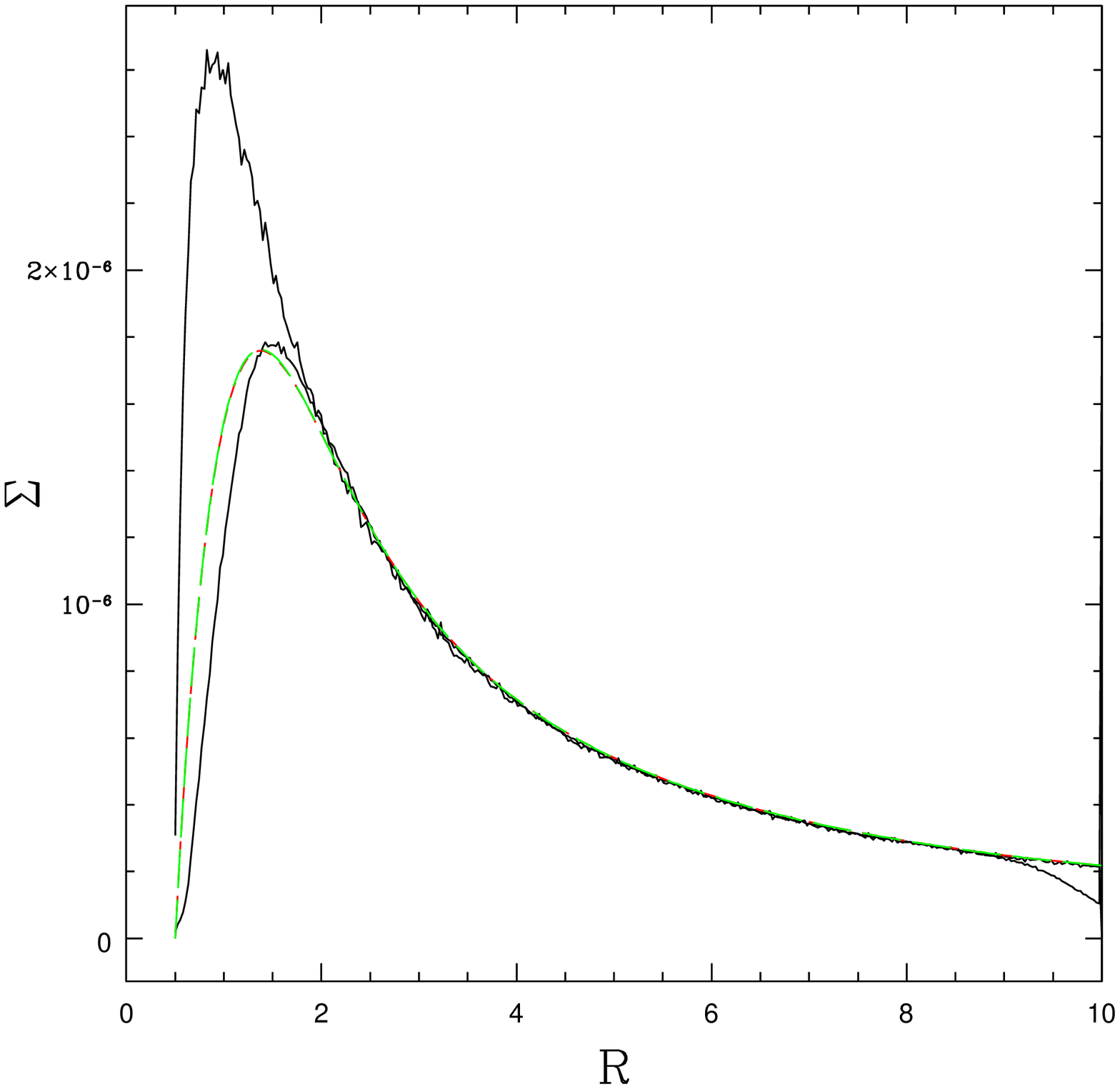}
\includegraphics[width=\columnwidth]{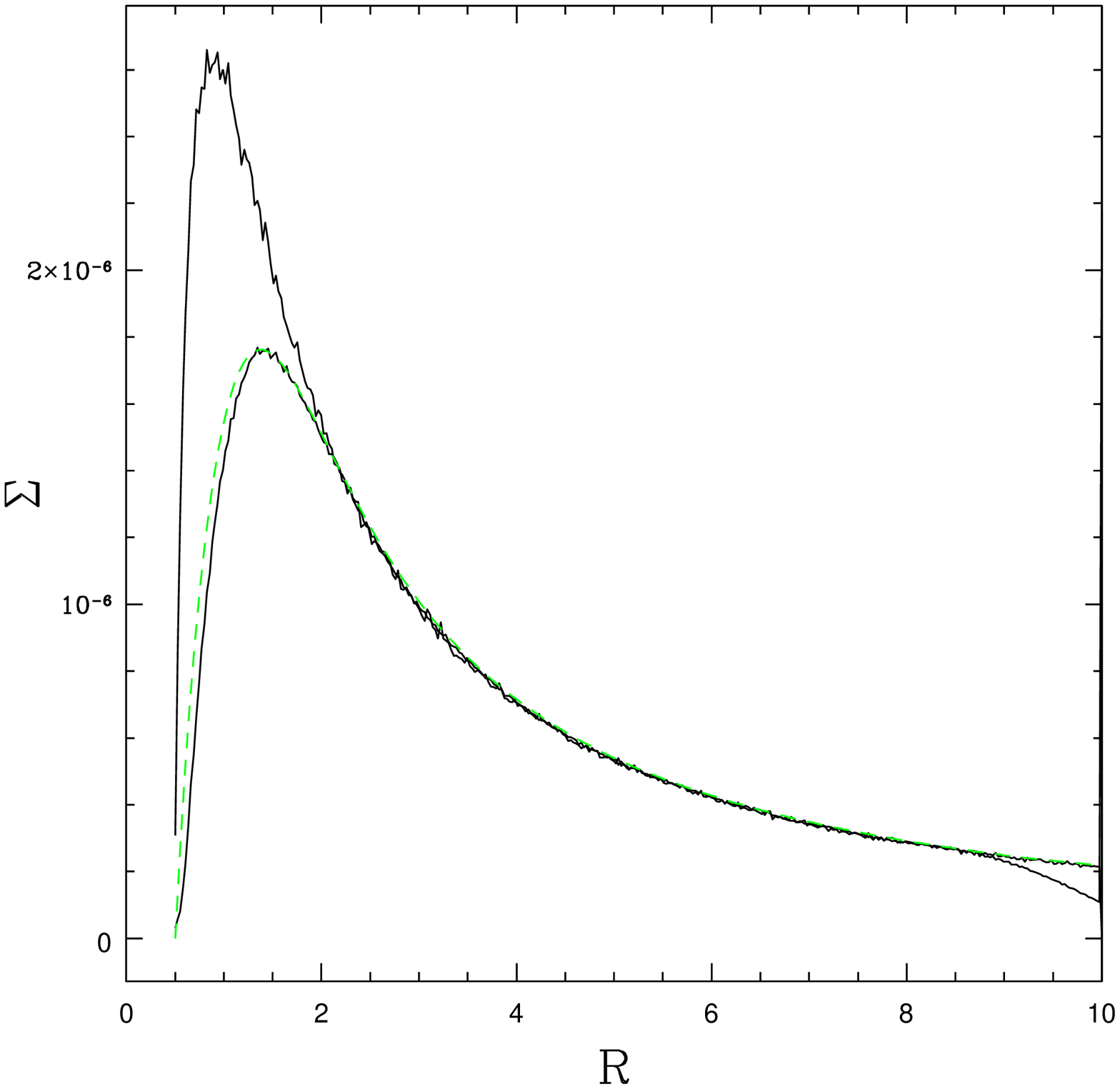}
\caption{Left: surface density profile from the SPH simulation (solid black line) and from the 1D evolution at $t=0$ and $t=500$ in code units, for the case $\alpha=0.3$, using the input value (from Eq. \ref{eq:nusph})  (dashed green line) and the measured best fit value (dashed red line) of the disc viscosity for the calculation using artificial viscosity to model disc viscosity at an SPH resolution of 2 million particles. We find very close agreement between the input and fitted values. Right: Same, but using Navier-Stokes viscosity. In this case, the fit with the $\Sigma$ profile is slightly better, giving smaller error bars in the fitted value of $\alpha$.}
\label{fig:fitalpha}
\end{center}
\end{figure*}

\subsection{Fitting procedure}
\label{sec:fit}
The main aim of this paper is to determine the relation between the two parameters $\alpha$ and $\alpha_2$, which describe the disc viscosity and the warp diffusion coefficient, respectively. In principle, the parameter $\alpha$ should be simply determined by the input viscosity coefficient in the SPH code as per Sec.~\ref{sec:sph}. However, \citetalias{LP07} did not find a perfect match between the nominal value of $\alpha$ as expected from the continuum limit of SPH and the $\alpha$ measured from the 1D surface density evolution using Eq. (\ref{eq:pringle}), and therefore preferred to fit both parameters independently, to get the desired relation. In particular, \citetalias{LP07} \emph{``stress that [they] do not perform an actual statistical fit of the viscosity coefficients, but simply choose them so as to match the evolution of the numerical simulation''}. This point is discussed further in Section \ref{sec:calib}. 

In this paper, we have implemented a statistical fitting procedure to check the calibration between the input $\alpha$ parameter and the value measured from comparing the SPH and the 1D evolution of the disc using \ref{eq:pringle}. The same procedure is further used to measure the warp diffusion coefficient $\alpha_2$ and the precession coefficient $\alpha_3$. This procedure is described below. Specific results for the calibration of $\alpha$ and the estimate of $\alpha_2$ and $\alpha_3$ are reported in Sec. \ref{sec:results}.

\subsubsection{Fittting for $\alpha$}

The viscosity parameter $\alpha$ is primarily responsible for the evolution of the disc surface density $\Sigma$. In particular, given the shape of the surface density in our initial condition, the feature which is most directly related to $\alpha$ is the decline of the peak surface density in the inner disc. In order to obtain $\alpha$ we have therefore fitted the shape of the surface density profile close to the peak as resulting from the 1D disc evolution at a given time to the SPH data. Thus, we have compared the data at $t=500$ (in code units), and considered an annulus of radial width equal to 0.1 each side of the maximum. To obtain $\alpha$, we have minimized the $L_2$ norm of the difference between the 1D evolution profile and the SPH data:
\begin{equation}
E_\alpha = \sum_i [\Sigma_i - \Sigma^{\rm 1D}(R_i)]^2,
\end{equation}
where the sum is taken over all shells (see previous section) within a radial distance 0.1 from the maximum. The value of $\Sigma^{\rm 1D}$ at $R_i$ is obtained by interpolation between the closest 1D cells. The minimum $E_{\alpha}$ is found by using a simple Newton-Raphson scheme. Starting from a trial value of $\alpha$ we iterate using:
\begin{eqnarray}
\alpha_{n+1} & = & \alpha_{n}  - \frac{E_\alpha^{'}(\alpha_{n})}{E_\alpha^{''}(\alpha_{n})} \\
 & = & \alpha_{n} - \frac{\epsilon}{2}\left[ \frac{E_\alpha(\alpha_{n}+\epsilon) - E_\alpha(\alpha_{n}-\epsilon)}
 {E_\alpha(\alpha_{n}+\epsilon) + E_\alpha(\alpha_{n}-\epsilon) - 2 E_\alpha(\alpha_{n}) }\right],\nonumber
\end{eqnarray}
where in the second line the first and the second derivatives of $E_\alpha$ are approximated by their finite difference value with respect to a small increment $\epsilon$. Once a minimum is found, we make sure that it is not a local minimum by checking that $E_\alpha$ is larger upon incrementing $\alpha$ by $5\epsilon$ either side of the minimum. We also compute the $1-\sigma$ uncertainty on the fitted value of $\alpha$ by computing the distance from the minimum at which $E_\alpha$ is increased by a factor 2.

An example of the best fit $\Sigma$ profile compared to the averaged SPH profile is shown in Fig. \ref{fig:fitalpha}. As it turns out, the best fit profile is in fact very close to the one computed using the input value for $\alpha$, so this fitting procedure for $\alpha$ itself becomes unnecessary (see Sec.~\ref{sec:calib}, below).

\subsubsection{Fittting for $\alpha_2$ and $\alpha_3$}

The values of $\alpha_2$ and $\alpha_3$ are obtained using a similar procedure, though the computation of either requires, and is dependent on, an input value for the disc viscosity $\alpha$ --- that is, using either the nominal or fitted $\alpha$ value.
% \emph{after the value of $\alpha$ has been fitted}, as described above. 

The diffusion coefficient $\alpha_2$ is mostly responsible for the evolution of the profile of ${\bf l}$, and in particular, given our initial conditions, it affects the evolution of $l_x$ around the warp radius at $R=R_0$. We therefore define the $L_2$ norm of the difference between the 1D evolution and the SPH evolution of $l_x$ at time $t=1000$ code units as:
\begin{equation}
E_{\alpha_2} = \sum_i [l_{i,x}- l_x^{\rm 1D}(R_i)]^2,
\end{equation}
where the sum is taken over all shells within a radial distance equal to 3 from the warp radius. In practice, rather than fitting $\alpha_2$ we fit the parameter $f$, defined as $\alpha_2 = f/(2\alpha)$. The parameter $f$ and its uncertainty are then obtained through minimization of $E_{\alpha_2}$ using a scheme analogous to the one used for $\alpha$.

The precession coefficient $\alpha_3$ affects primarily the evolution of $l_y$ around the warp radius. Its best fit value is thus obtained from the minimization of
\begin{equation}
E_{\alpha_3} = \sum_i [l_{i,y}- l_y^{\rm 1D}(R_i)]^2,
\end{equation}
using a scheme analogous to the one used for $\alpha$ and $\alpha_2$, and depending on the input values of both of these. As for $\alpha_{2}$ the sum is taken over all shells within a radial distance of 3 from the warp radius.

\section{Results}
\label{sec:results}

The initial aim of this paper was to perform simulations at a resolution significantly higher than that employed by \citetalias{LP07}, in order to assess the effect of limited resolution. Having performed several calculations with 20 million SPH particles and finding results indistinguishable from the lower resolution of 2 million particles, we have instead surveyed a wide range in parameter space, performing a total of 78 simulations using 2 million particles together with the original 8 at 20 million particles. These consist of 8 series of simulations, each for a range of viscosity values. The parameters for each series are given in Table~\ref{tab:runs}, where we have considered the effect of resolution (2 vs. 20 million particles), three different warp amplitudes ($A=0.01$, $0.05$ and $0.5$), disc viscosity formulated either using the modified artificial viscosity (AV) or by a direct implementation of Navier Stokes terms (NS), considering the latter with zero bulk viscosity and subsequently with a small amount applied using a switch.

\begin{table}
\begin{tabular}{lp{0.7cm}p{0.7cm}p{0.7cm}ll}
\hline
Series & A & Visc. type & Bulk visc.? & N$_{\rm part}$ & Symbols\\
\hline
1 & 0.01 & AV & yes & $2 \times 10^{6}$ & red triangles \\
2 & 0.05 & AV & yes & $2 \times 10^{6}$ & orange triangles \\
3 & 0.01 & AV & yes & $2 \times 10^{7}$ & green triangles \\
4 & 0.05 & AV & yes & $2 \times 10^{7}$ & cyan triangles \\
5 & 0.01 & NS & no & $2 \times 10^{6}$ & black squares \\
6 & 0.01 & NS & switch & $2 \times 10^{6}$ & blue squares \\
7 & 0.5 & AV & yes & $2 \times 10^{6}$ & magenta triangles \\
8 & 0.5 & AV & yes & $2 \times 10^{7}$ & yellow triangles \\
\hline
\end{tabular}
\caption{Parameter settings for each of the 8 series of calculations performed in this paper, where each series consists of a set of simulations covering a range of disc viscosities $\alpha$. The second column gives the initial warp amplitude $A$ used in Eq.~(\ref{eq:setup}), the third column shows whether the disc viscosity was represented using the modified artificial viscosity term (AV) or via a direct implementation of Navier-Stokes viscosity (NS). The fourth column shows whether or not bulk viscosity was applied (always true for the AV calculations). The resolution of the calculations is given in the fifth column and the symbols used to represent each series in Figures~\ref{fig:alphavsalpha}, \ref{fig:alpha2}, \ref{fig:flebbe}, \ref{fig:largeA} and \ref{fig:alpha3} are given in the last column.}
\label{tab:runs}
\end{table}

\subsection{Calibration of the disc viscosity coefficient}
\label{sec:calib}

 \begin{figure}
\begin{center}
\includegraphics[width=\columnwidth]{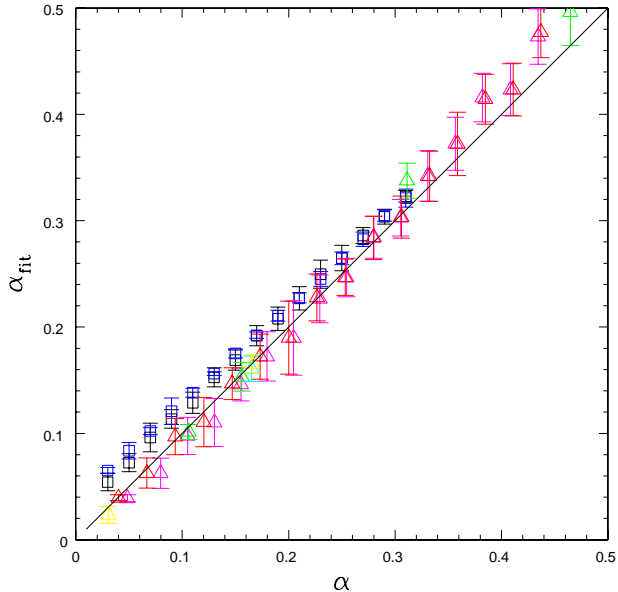}
\caption{Comparison between the input value of $\alpha$ and the measured value obtained fitting the SPH data to the 1D evolution of $\Sigma$, with the zero torque boundary condition enforced exactly. Results are shown with triangles for the calculations where the disc viscosity has been simulated using the SPH artificial viscosity, whilst squares correspond to calculations where physical viscosity terms have been implemented directly. All calculations are performed at a resolution of 2 million SPH particles, except for the green, cyan and yellow triangles that use 20 million particles. Error bars refer to the 1$\sigma$ errors from the fitting procedure used to obtain $\alpha_{\rm fit}$. }
\label{fig:alphavsalpha}
\end{center}
\end{figure}

The best matching values of $\alpha$ (here referred to as $\alpha_{\rm fit}$) reported in Table 1 of \citetalias{LP07} do not follow the expected relation given by Eq. (\ref{eq:nusph}).  Although, for a given $\langle h\rangle/H$, the relation between $\alpha_{\rm fit}$ and $\alpha^{\rm AV}$ is approximately linear, the slope is shallower than expected. An accurate calibration of $\alpha$ is important, because in turn it is used as an input for the evaluation and fitting of the warp diffusion coefficient $\alpha_2$. The disagreement found in \citetalias{LP07} prompted us to examine the method used to calibrate $\alpha$ in greater detail, resulting in our implementation of the fitting procedure described in Section \ref{sec:fit} --- essentially a quantitative version of the procedure performed in \citetalias{LP07}. 

In considering this issue, we have also explored the effect of the inner boundary condition of the 1D disc evolution on the measurement of $\alpha$. Indeed, the main feature which is used for the evaluation of $\alpha$ is the turnover of the surface density at small radii, which might well be affected by the specific boundary condition used. Using the same condition employed by \citetalias {LP07} (described in Sec. \ref{sec:discevol}, above), we found a similar relationship between $\alpha_{\rm fit}$ and $\alpha^{\rm AV}$ --- that is, not matching the expectations from Eq. (\ref{eq:nusph}). Furthermore, we found that the calculations using a Navier-Stokes viscosity also did not agree with the measured values. However, when the zero torque boundary condition was enforced exactly (see Sec. \ref{sec:discevol}), the disagreement was removed. This is demonstrated in Fig. \ref{fig:alphavsalpha}, which shows the best fit value $\alpha_{\rm fit}$ versus the input value of $\alpha$ calculated using Eq. (\ref{eq:nusph}). Results are shown with triangles for the calculations where the disc viscosity has been simulated using the SPH artificial viscosity, whilst squares correspond to calculations where Navier-Stokes viscosity terms have been implemented directly. All calculations are performed at a resolution of 2 million SPH particles, except for the green and cyan triangles that use 20 million particles. The error bars show the 1$\sigma$ errors from the fitting procedure.

As one can see from Fig. \ref{fig:alphavsalpha}, the general agreement is very good. In other words, the procedure used by \citetalias{LP07} to simultaneously fit $\alpha$ and $\alpha_2$ is no longer necessary and henceforth we simply fit the single parameter $\alpha_2$ using the input value for $\alpha$. 

Also worth noting from Fig. \ref{fig:alphavsalpha} is that, whilst the error bars ar smaller using the Navier-Stokes viscosity (squares) --- due to an improved agreement of the profile of $\Sigma$ near the inner boundary (see right panel of Fig. \ref{fig:fitalpha}) ---, there appears to be a small excess dissipation that occurs for low input $\alpha$. In the case of Series~5 (black squares), this may be naturally attributed to the small amount of artificial viscosity we have applied. However, for Series~6 (blue squares), using zero artificial viscosity paradoxically results in an even \emph{greater} dissipation. We attribute this to the increased randomness of the particle distribution arising when no bulk viscosity is present. A similar effect is seen in several other SPH simulations, e.g. in \citet{pf10} when $\beta^{\rm AV}$ is too low.

\subsection{Results for the warp diffusion coefficient}

 \begin{figure*}
\begin{center}
\includegraphics[width=1.4\columnwidth]{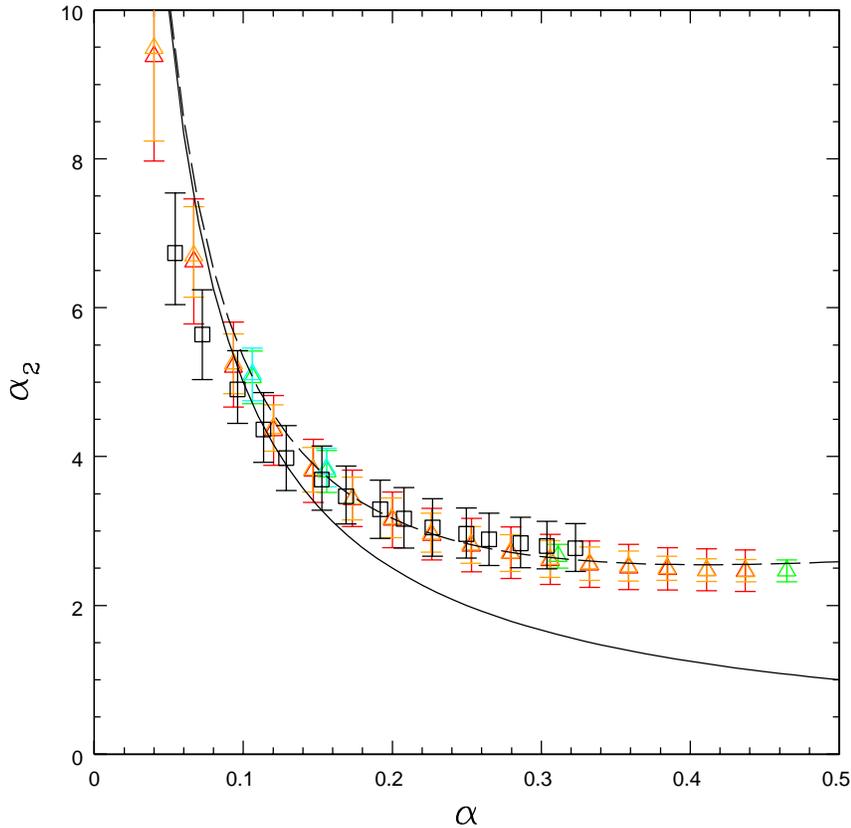}
\caption{Relation between the warp diffusion coefficient $\alpha_2$ and the disc viscosity $\alpha$, for warp amplitudes $A=0.01$ (black squares, red and green triangles) and $A=0.05$ (orange and cyan triangles). Squares use a direct implementation of the Navier-Stokes terms, whilst triangles use the SPH artificial viscosity to represent the disc viscosity. All calculations employ 2 million SPH particles, except the cyan and green triangles, which use 20 million. The solid line shows the simple $\alpha_2=1/(2\alpha)$ relation from \citet{pappringle83}. The long-dashed line includes the non-linear corrections due to finite values of $\alpha$ for small amplitude warps (Eq. \ref{eq:prediction2}).}
\label{fig:alpha2}
\end{center}
\end{figure*}

The results of the fitting procedure for the warp diffusion coefficient $\alpha_2$ are shown in Fig. \ref{fig:alpha2}, for Series 1 to 5 discussed above. The solid line in the figure shows the simple relation $\alpha_2=1/(2\alpha)$ expected in the linear regime of warp propagation discussed in Section \ref{sec:theory} \citep{pappringle83}. What is most striking is that the numerical results do not appear to match the $1/(2\alpha)$ relation in almost \emph{any} regime. In particular, at high $\alpha$, the measured values of $\alpha_2$ lie much above the simple $1/(2\alpha)$ relation, while at low $\alpha$ they lie slightly below it. This is contrary to the result shown in \citetalias{LP07}, where agreement was found at high $\alpha$, and a much larger disagreement was found at low $\alpha$. The origin of the discrepancy between our results and those of  \citetalias{LP07} lies in the more accurate evaluation of the disc viscosity $\alpha$ discussed above and the resultant effect on the fit for $\alpha_2$. 

The natural question is therefore: why do the new, more accurate results not agree with the standard theory?

Initially, we will discuss only the simulations with a small warp amplitude $A=0.01$. The effect of finite warp amplitude is discussed later in Section \ref{sec:amplitude}

\subsubsection{Effect of resolution}

Our first attempt to reconcile the simulation with the theory was to check for resolution effects. In particular, these are thin discs, and the vertical scale-height is only moderately resolved (by only 1.6 smoothing lengths) at a resolution of 2 million particles (see Fig.~\ref{fig:xsection}). \citetalias{LP07} also mention the possibility of resolution effects in their simulations (which have the same resolution of 2 million particles) but nevertheless demonstrate that the vertical density profile is very well reproduced (see Fig. 2 of \citetalias{LP07}), and hence argue that such effects should be negligible. We have therefore performed 6 calculations at the higher resolution of 20 million particles (four with $A=0.01$ and two with $A=0.05$), for which the vertical scale-height is resolved with $\sim$ 3.5 smoothing lengths. The results of these higher resolution simulations are shown with green and cyan triangles in Fig. \ref{fig:alpha2} and show no significant difference with respect to the lower resolution case. Note that the calibration of $\alpha$ for these calculations, shown in Fig. \ref{fig:alphavsalpha} is very similar to the 2 million particle case, despite the very different smoothing lengths. We therefore conclude that resolution effects are not to blame for the discrepancy.

\subsubsection{Effect of viscosity formulation}
   
The results of \citetalias{LP07} showed strong disagreement with the standard theory at low $\alpha$. \citetalias{LP07} argue that a discrepancy between their results and the standard theory at low $\alpha$ might arise because of enhanced dissipation due to the presence of supersonic motions and hence shocks. With the recalibration of $\alpha$ discussed above the disagreement at low $\alpha$ is significantly reduced and, as we show later, can now be explained by the transition to the wave-like propagation regime. 

Nevertheless, when using the SPH artificial viscosity to represent disc viscosity, one inevitably ends up with a large bulk viscosity coefficient (see Section \ref{sec:avvisc}, Eq. \ref{eq:bulkav}), which could perhaps explain the excess dissipation invoked by \citetalias{LP07}. In order to check whether bulk viscosity affects the magnitude of the warp diffusion coefficient, we have implemented a full Navier-Stokes viscosity, as described in Section \ref{sec:numerics}. 

\begin{figure}
\begin{center}
\includegraphics[width=\columnwidth]{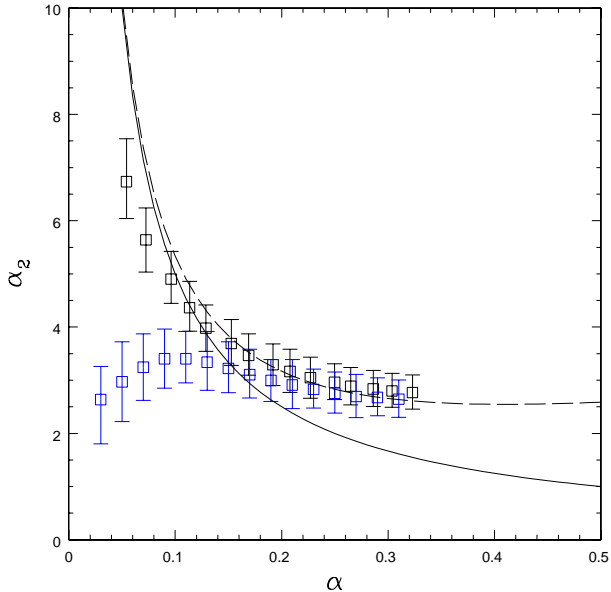}
\caption{As in Fig. \ref{fig:alpha2}, but comparing only the two series using Navier-Stokes viscosity, with (black suqares) and without (blue squares) a small amount of artificial bulk viscosity. Despite the nominally lower input viscosity, the calculations with zero bulk viscosity paradoxically show a larger dissipation, as evidenced by the lower magnitude of the warp diffusion coefficient $\alpha_2$. We attribute this to the danger of using zero bulk viscosity in SPH.}
\label{fig:flebbe}
\end{center}
\end{figure}

The results of the calculations using the Navier-Stokes viscosity and zero bulk viscosity and the artificial viscosity set to zero are shown in Fig. \ref{fig:flebbe} (blue squares), and indeed appear to show a significant effect, lowering the measured values of $\alpha_2$  at low $\alpha$. But, as discussed in great detail in \citetalias{LP07}, this implies \emph{an even greater excess dissipation}. Thus, paradoxically, by removing bulk viscosity we have apparently \emph{increased} the amount of dissipation present in the simulation. However, some caution is required here, since --- as discussed in section \ref{sec:flebbe} --- one should be very careful about completely removing all bulk viscosity from SPH simulations, since it plays a necessary role in preventing particle interpenetration, as well as providing physical dissipation in shocks, if shocks are present. 

We have therefore also computed a series of simulations with Navier-Stokes shear viscosity, but with a small amount of artificial viscosity, applied only for approaching particles and using a switch to reduce its amplitude away from shocks (black squares in Figs.~\ref{fig:alpha2} and \ref{fig:flebbe}). Note that the effective bulk viscosity coefficient in this case is much smaller (at least a factor of 5) that that present in the simulations that use artificial viscosity to represent disc viscosity (i.e. the red triangles in Fig.~\ref{fig:alpha2}). Despite the very small change in the viscosity formulation, the results of this series of simulations (series 5, black squares in Fig.~\ref{fig:alpha2}) are in very good agreement with those presented previously (series 1, red triangles). This leads us to conclude that the dramatic effect produced in Fig.~\ref{fig:flebbe} by removing all of the bulk viscosity, is most likely a numerical artefact. This is strengthened by the results already discussed in Fig. \ref{fig:alphavsalpha}, also showing that the simulations with zero bulk viscosity show a higher dissipation, that can be attributed to the increased randomness of the particle distribution when no bulk viscosity is present. 

\subsubsection{Are we looking at the right theory?}

Having investigated the effect of both resolution and viscosity formulation, we may dispense with the possibility that the disagreement between the SPH simulations and the simple linear theory is simply a numerical artefact. What is worse, the recalibration of $\alpha$ discussed in Section \ref{sec:calib} puts us in a worse position than \citetalias{LP07}, as we now disagree with the linear theory at both low \emph{and high} values of $\alpha$, contrary to the results of \citetalias{LP07}, where the results appeared to agree at high $\alpha$. In our case, the results are also more significant, because we have much better statistics (86 simulations compared to the 10 shown in Fig. 8 of \citetalias{LP07}), spanning a wider range in all of the parameters $\alpha$, $\alpha_2$ and the warp amplitude. 

Clearly, though, the $\alpha$ values we are using here at not small, and therefore it may well be expected that the nominal  $1/(2\alpha)$ relation is not appropriate, particularly at the high-$\alpha$ end of our parameter range. Indeed, when we plot 
the relation between $\alpha_2$ and $\alpha$ predicted for finite values of $\alpha$ by Eq. (\ref{eq:prediction2}) 
(long-dashed line in Figs.~\ref{fig:alpha2} and \ref{fig:flebbe}), we recover an excellent agreement with the SPH results for $\alpha\gtrsim 0.15$ for low-amplitude warps. 

We thus conclude that there is no disagreement between our numerical results at high $\alpha$ and the linear theory of warp propagation, once the effects of finite $\alpha$ are appropriately accounted for in the theory.  The only remaining issue is a small disagreement at low $\alpha$ for small amplitude warps --- though much smaller than the one found in \citetalias{LP07}, which we discuss in the next section. 

\subsubsection{Low viscosity behaviour and transition to the wave-like propagation regime}
\label{sec:wave}

\begin{figure}
\begin{center}
\includegraphics[width=\columnwidth]{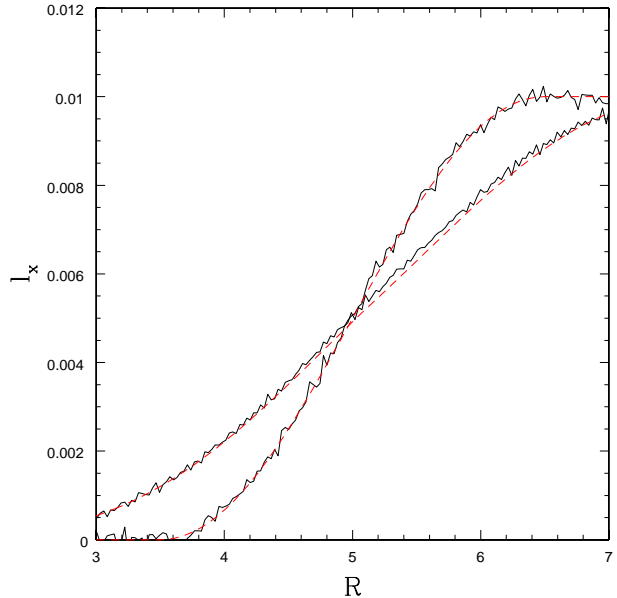}
\caption{Evolution of the shell averaged profiles of $l_x$ from the SPH simulations at $t=0$ and $t=500$ in code units (solid black lines), for the low viscosity case $\alpha=0.065$ and $A=0.01$, compared to the results of a 1D evolution assuming wave-like propagation with dissipation at the appropriate value of $\alpha$, from Eqs. (\ref{eq:wave1}) and (\ref{eq:wave2}) (dashed red lines). The excellent agreement explains the deviations at low $\alpha$ seen in Fig.~\ref{fig:alpha2}, as been due to the transition to the wave-like propagation regime at low viscosity.}
\label{fig:wave}
\end{center}
\end{figure}

In principle, there could be two explanations for the remaining small disagreement that we find for low viscosity: a numerical one, related to our procedure to fit the value of the diffusion coefficient $\alpha_2$, and a physical one, related to an actual transition to a different propagation regime, such as the wave-like propagation regime (sec. \ref{sec:theory}). 

There are good reasons to believe that both effects might play a role for small values of the viscosity coefficient. From the numerical point of view, we note that our fitting procedure is based on matching the solution of the simple diffusion equation to the SPH results at a given time, $t=1000$ in code units, which corresponds to $0.4\alpha$ in units of the viscous time at $R=1$. The viscous time is not only a measurement of the time needed for the overall viscous evolution of the disc to take place, but also of the time needed to smooth out the discreteness of the particle distribution in the initial condition. We therefore might expect that, for small $\alpha$, the SPH simulation at $t=1000$ code units is still somewhat noisy, and that it might affect the evaluation of $\alpha_2$. That this is the case can already be seen from the fact that the error bars on $\alpha_2$ get larger at small $\alpha$. Additionally, we have also found that for small $\alpha$ the results are somewhat sensitive to the time at which the fit is performed. If the fit is performed at earlier times, the resulting value of $\alpha_2$ tends to be systematically shifted down by a small amount. 

The results of the calculation for low $\alpha$ might also be affected by the fact that warp propagation undergoes a transition to the wave-like regime (sec. \ref{sec:theory}), which, for small warp amplitudes, is expected to occur at around $\alpha\lesssim 2\pi H/\lambda\approx 0.07$, where $\lambda$ is the wavelength of the perturbation (which in our case is $\lambda\sim 6$ in code units). In order to test this hypothesis we compare the evolution of the SPH simulation to the 1D evolution appropriate to the wave-like regime (Eqs. \ref{eq:wave1} and \ref{eq:wave2}), including dissipation corresponding to the input value of $\alpha$ from the simulation. The results of this test are shown in Fig. \ref{fig:wave} for a low amplitude warp with $\alpha= 0.065$, from Series 1. The plot shows the profile of the $x$-component of the unit vector ${\bf l}$ from the SPH simulation at $t=0$ and $t=500$ in code units (solid black lines) compared to the 1D evolution from Eqs. (\ref{eq:wave1}) and (\ref{eq:wave2}) (dashed red lines). The fact that the profiles agree well indicates that wave-like propagation, albeit with diffusion, can explain the deviation from the purely diffusive propagation we have assumed when comparing with Eq. (\ref{eq:prediction}) or Eq. (\ref{eq:prediction2}).

Given that there is no longer any disagreement between theory and the results of the SPH simulations, contrary to \citetalias{LP07}, it is no longer necessary to invoke any additional dissipation involved in warp propagation.

\subsubsection{Large amplitude warps}
\label{sec:amplitude}

\begin{figure}
\begin{center}
\includegraphics[width=\columnwidth]{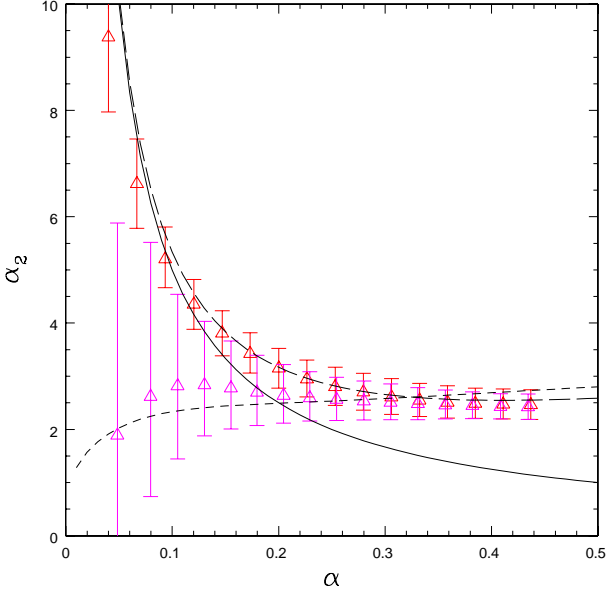}
\caption{Relation between the warp diffusion coefficient $\alpha_2$ and $\alpha$ for large amplitude warps. The magenta triangles show the results for $A=0.5$, while the red triangles, for comparison, indicate the small amplitude ($A=0.01$) case. The solid and long-dashed lines show the $1/(2\alpha)$ relation and the corrections expected for finite $\alpha$, respectively. The short-dashed line shows the expected relation based on the non-linear theory of \citetalias{ogilvie99}, assuming a fixed  $\psi=0.55$ (roughly comparable to the average $\psi$ value using $A=0.5$). The large error bars in the high amplitude case at low $\alpha$ are because the non-linear warp evolution is not well fitted by a single value for $\alpha_{2}$ --- physically manifested as a steepening of the warp profile observed in the simulations (see Fig.~\ref{fig:profilelargeAsmallalpha}) that is not captured by the fitted linear profiles.}
\label{fig:largeA}
\end{center}
\end{figure}

\begin{figure}
\begin{center}
\includegraphics[width=\columnwidth]{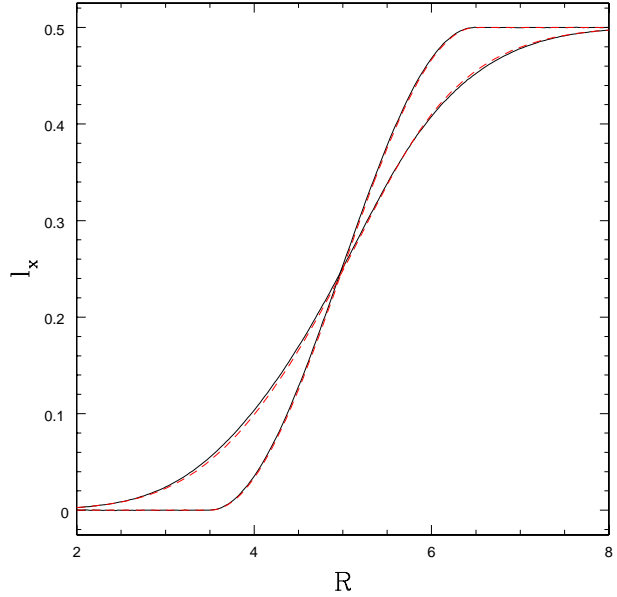}
\caption{Shell averaged profiles of $l_x$ from the SPH simulation (solid black lines) at $t=0$ and $t=1000$ (in code units), compared to the corresponding profiles from the diffusive evolution, for the high viscosity case $\alpha=0.43$ and a strongly non-linear warp amplitude ($A=0.5$), assuming the warp diffusion coefficient to be a function of $\alpha$ and $\psi$, as predicted by \citetalias{ogilvie99} (dashed red lines).}
\label{fig:profilelargeA}
\end{center}
\end{figure}

Having gained some confidence that the linear theory of warp propagation explains satisfactorily the evolution of the SPH simulations at low warp amplitudes, we may turn our attention to non-linear effects. 

Initially we have considered a small increase in the initial warp amplitude to $A=0.05$ in Eq. (\ref{eq:setup}). The results of these calculations (Series 2) are shown with orange triangles in Fig.~\ref{fig:alpha2} and are essentially indistinguishable from the $A=0.01$ case (Series 1 and 5), demonstrating that the linear theory is applicable also in this case.

The resulting values of $\alpha_2$ from our fitting procedure for the $A=0.5$ case (Series 7) are shown in Fig. \ref{fig:largeA} with magenta triangles, compared to the corresponding Series 1 for $A=0.01$ (red triangles). One can immediately see that the fitted values of $\alpha_2$, for small $\alpha$, are much smaller than the low amplitude case, and are characterized by a much larger uncertainty. 

At very large warp amplitudes, the linear theory of warp propagation in the diffusive regime is not applicable, and it is therefore not surprising that, indeed, the SPH simulations in this regime ($A=0.5$) do not match either the $1/(2\alpha)$ relation or the more complete relation of Eq. (\ref{eq:prediction2}) which is non-linear in $\alpha$ but assumes linearity in the warp amplitude. \citetalias{ogilvie99} presents a non-linear theory of warp propagation for warps of any amplitude. We are now in a position to test this theory numerically.

The complicating factor is that, in the \citetalias{ogilvie99} theory, $\alpha_2$ is a function of the warp amplitude $\psi$ (related to $A$ by Eq. \ref{eq:psi}, with the maximum value given by Eq. \ref{eq:psimax}). We thus cannot associate a single value of $\alpha_2$ to each large amplitude simulation, as $\psi$ is a function of radius (Eq. \ref{eq:psi}) and furthermore decreases as a function of time as the warp diffuses. Indeed, this is the reason for the large error bars in Fig.~\ref{fig:largeA} when attempting to fit a single $\alpha_{2}$ value to the $l_{x}$ profile, particularly at low $\alpha$ where in the \citetalias{ogilvie99} theory the dependency of $\alpha_{2}$ on $\psi$ is much stronger. What is possible is to check whether the deviations expected from such non-linear theory follow the observed trend, i.e. a general decrease value of the diffusion coefficient. The short-dashed line in Fig. \ref{fig:largeA} shows the relation between $\alpha_2$ and $\alpha$ based on the theory of  \citetalias{ogilvie99} for a fixed value of $\psi=0.55$, computed numerically based on a routine kindly provided to us by Gordon Ogilvie. We can indeed see that the non-linear theory does reproduce qualitatively the observed trend.

As an additional test of the theory, we have also evolved the standard equation for diffusive evolution (Eq. \ref{eq:pringle}), where the diffusion coefficients $\alpha_{1}$ (now different from $\alpha$) and $\alpha_2$ are computed at each radius and at each timestep, based on Ogilvie's non-linear theory. In this case, we do not have any free parameter to fit: $\alpha$ is the nominal value of the viscosity coefficient, and $\alpha_{1}$ and $\alpha_2$ are prescribed functions of $\alpha$ and $\psi$ which we compute at each radius and time by evaluation of the relevant integrals in \citetalias{ogilvie99} using the routine provided. The resulting profiles of $l_x$ are shown in Fig. \ref{fig:profilelargeA} for the case where $\alpha=0.43$. The black solid lines show the results of the SPH simulations at $t=0$ and $t=1000$ (in code units), while the red dashed line show the corresponding profiles computed from Eq.  (\ref{eq:pringle}). The agreement of the non-linear theory with the SPH results is excellent. Note that the warp diffusion coefficient in the theory of \citetalias{ogilvie99} depends also on the amount of bulk viscosity present in the disc. Indeed, in order to get the good match shown in Fig. \ref{fig:profilelargeA} we have computed the warp diffusion coefficient assuming a bulk viscosity with magnitude $\alpha_{\rm b}=5\alpha/3$, as expected from the SPH formalism (see sect. \ref{sec:av}).

\begin{figure}
\begin{center}
\includegraphics[width=\columnwidth]{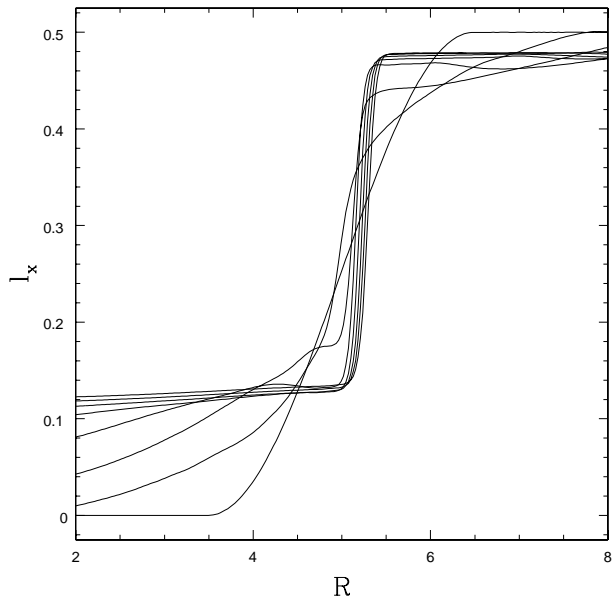}
\caption{Shell averaged profiles of $l_x$ from the SPH simulation (black lines) at intervals of $\Delta t=500$ up to $t=3500$ (in code units), for the low viscosity case $\alpha=0.03$, using 20 million particles and a strongly non-linear warp amplitude ($A=0.5$). The large warp amplitude simulations at low $\alpha$ show a strong steepening effect in the warp profile because different parts of the warp propagate at different speeds. A comparison with the non-linear theory is more difficult in this case because the predicted diffusion parameter for the disc can become negative, causing unphysical oscillations in the one-dimensional code.}
\label{fig:profilelargeAsmallalpha}
\end{center}
\end{figure}

\begin{figure}
\begin{center}
\includegraphics[angle=270,width=\columnwidth]{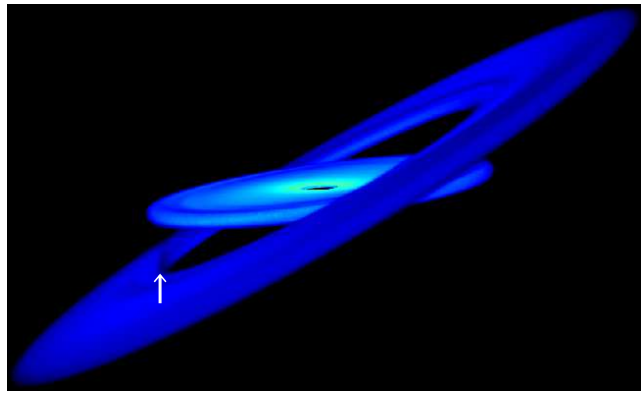}
\caption{Resulting 3D disc structure from the simulation shown in Fig.~\ref{fig:profilelargeAsmallalpha} with a large amplitude warp in a low viscosity disc ($\alpha = 0.03$), shown at $t=1500$ (in code units). The steepening of the warp profile in this case results in a nearly complete break in the disc. At the employed resolution of 20 million particles we are able to resolve the thin but steady stream of material that is nevertheless still being transported inwards across the discontinuity (just visible in the Figure, as indicated by the arrow).}
\label{fig:discbreak}
\end{center}
\end{figure}

 At low disc viscosities (small $\alpha$), the simulations show a steepening effect in the warp profile --- the physical result of different parts of the warp propagating at different speeds. We find that the steepening becomes stronger as the disc viscosity is reduced (evident from the increase in the error bars seen in Fig.~\ref{fig:largeA} as $\alpha \to 0$) and in the limit of extremely small viscosity can result in a near-complete break in the disc. This is demonstrated in Fig.~\ref{fig:profilelargeAsmallalpha} which shows the equivalent of Fig.~\ref{fig:profilelargeA} for a low viscosity calculation ($\alpha=0.03$) shown at intervals of $t=500$ in code units and evolved as far as $t=3500$. A three dimensional rendering of the disc structure is shown in Fig.~\ref{fig:discbreak} at $t=1500$ in code units. Despite the apparent break the calculations are able to resolve a thin but steady stream of material continuing to accrete across the discontinuity, indicating that the disc regions remain connected even in this extreme case.
 
 The simulation results in this case are more difficult to compare with theory because of the development of large unphysical oscillations around the warp in the one dimensional code. This is not unexpected from the non-linear theory, since in this regime the evolution of the disc is not well described by a diffusion equation. In particular, \citetalias{ogilvie99} (Eq. 141) shows that 
 \begin{equation}
 \alpha_1 = \alpha - \frac{1}{24}\frac{\psi^2}{\alpha},
 \end{equation}
 and thus for large $\psi$ (more specifically, when $\psi>\sqrt{24}\alpha$, which is the case here) the effective viscosity becomes negative, suggesting that diffusion is no longer an appropriate description. Whilst for small amplitude warps one expects a transition to the wave-like regime at low $\alpha$ (based on the dispersion relation derived from the linear wave equations, Eqs. (\ref{eq:wave1})-(\ref{eq:wave2})), it is not clear from theory at which value of $\alpha$ the transition occurs for large amplitude warps, and indeed whether such a transition occurs at all if $\psi$ is sufficiently large (Ogilvie, private communication). Furthermore, calculations by \citet{ogilvie06} based on the evolution of a one-dimensional non-linear Schr\"odinger equation for inviscid, Keplerian discs suggest that wave-like propagation should not have a steepening effect for an isothermal equation of state (more specifically, if the adiabatic index $\gamma < 3$), suggesting that the steepening we observe is indeed a diffusive rather than wave-like effect, though it is hard to be certain in the absence of a full non-linear theory for wave-like propagation.
 
\subsection{Precession}

\begin{figure*}
\begin{center}
\includegraphics[width=\columnwidth]{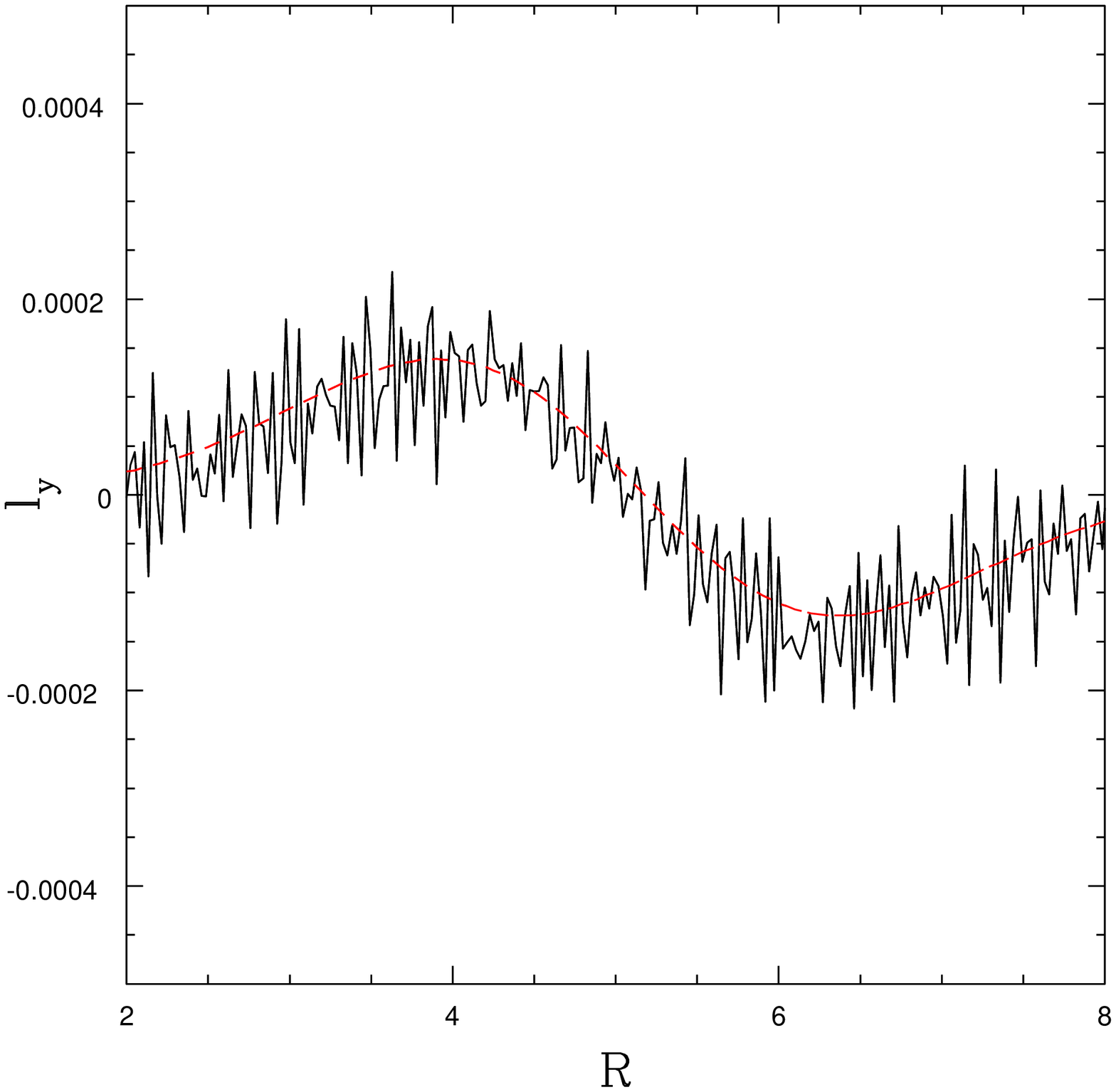}
\includegraphics[width=\columnwidth]{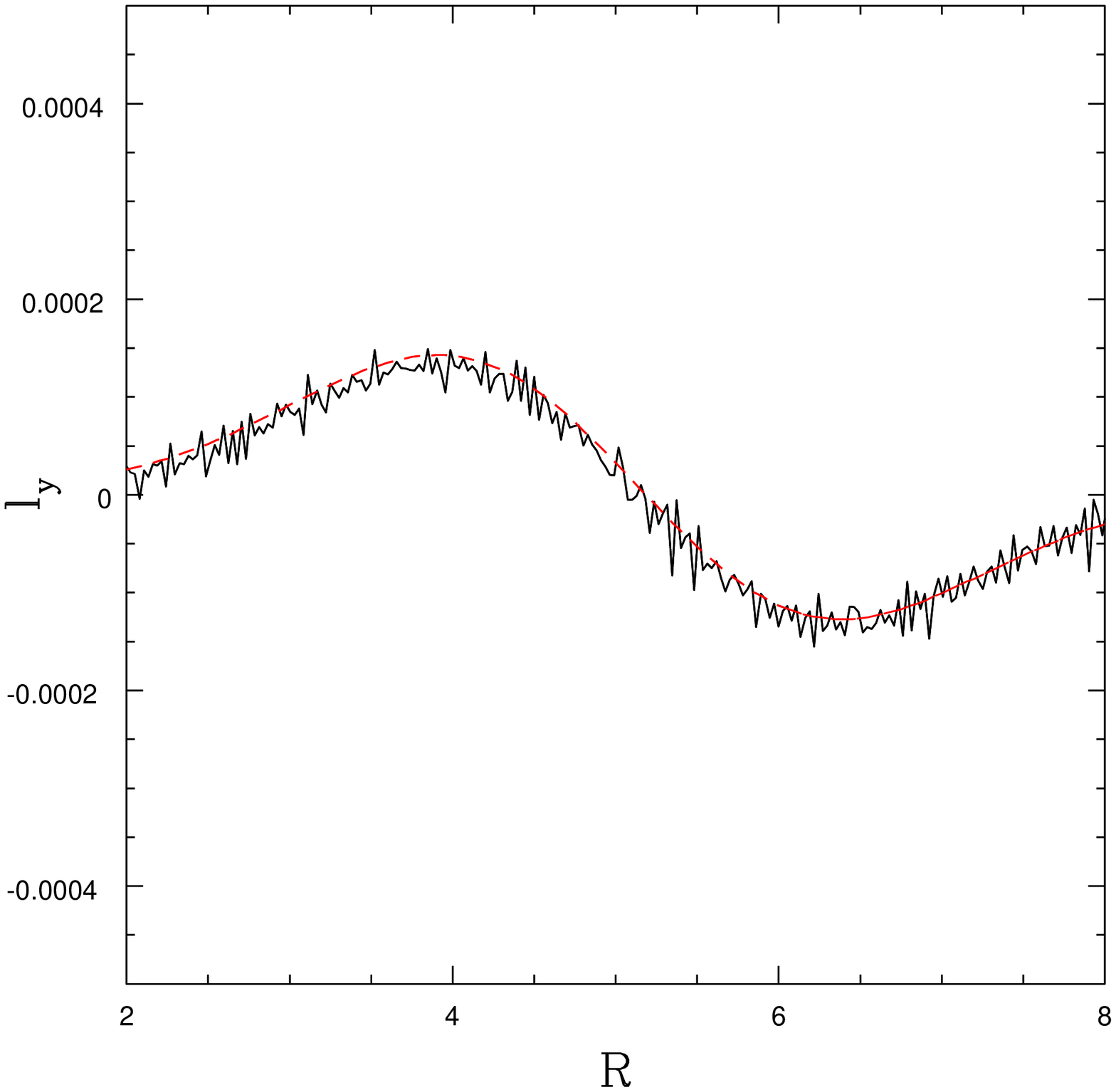}
\caption{Left: Profile of $l_y$ at $t=1000$ (in code units) for the $\alpha=0.43$ case. The solid black line shows the SPH results at a resolution of 2 million particles. The dashed red line shows the expected profile from the diffusion plus precession equation with the fitted value of $\alpha_3$. Right: Profile of $l_y$ at $t=1000$ code units for the $\alpha=0.46$ case, at an SPH resolution of 20 million particles (lines are the same as for the left panel).}
\label{fig:precession}
\end{center}
\end{figure*}

The last remaining piece of the theory which needs to be tested is the one related to precession. As mentioned in sect. \ref{sec:theory}, the full theory of warp propagation presented in \citetalias{ogilvie99} predicts also the presence of internal precessional torques. These can be accounted for within the simple diffusion model of \citet{pringle92} (Eq. \ref{eq:pringle}) by adding an appropriate additional term (Eq. \ref{eq:precession}). We have included such terms and fitted the value of the corresponding new parameter $\alpha_3$ to the SPH data. In this case, the disc property which relates to precession and that we use to perform the fit is the profile of the $y$-component of the angular momentum in the disc $l_y$. 

Figure \ref{fig:precession} shows the profiles of $l_y$ at $t=1000$ code units (solid black lines) and the corresponding solution of the diffusion equation with added precessional term, using the fitted value of $\alpha_3$ (dashed red lines). The left panel refers to $\alpha=0.43$ and an SPH resolution of 2 million particles, while the right plot refers to $\alpha=0.46$ and an SPH resolution of 20 million particles. Note that the amplitude of the induced precession --- and therefore of the value of $l_y$ --- is small, and therefore the SPH data are quite noisy for $l_y$, especially at low SPH resolution. This implies a rather large uncertainty in the fitted value of $\alpha_{3}$. 

\begin{figure}
\includegraphics[width=1.\columnwidth]{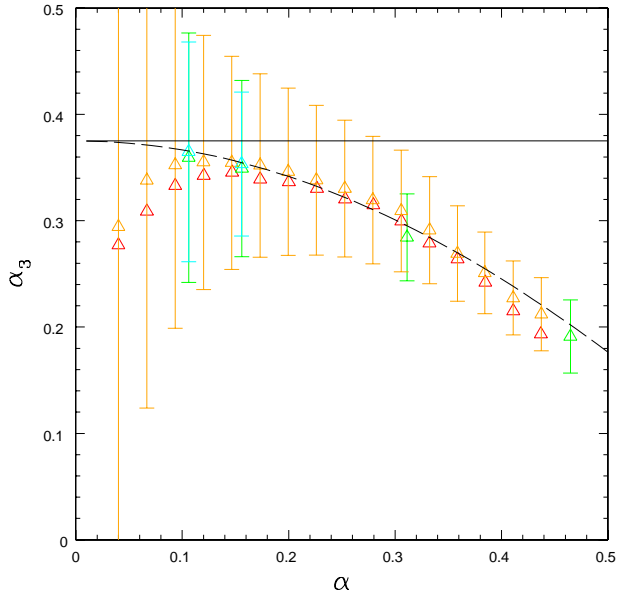}
\caption{Relation between the precession coefficient $\alpha_3$ and the disc viscosity $\alpha$, for warp amplitudes $A=0.01$ (red and green triangles) and $A=0.05$ (orange and cyan triangles). All calculations employ 2 million SPH particles, except the cyan and green triangles, which use 20 million. The solid line shows the expected precession rate in the limit of small $\alpha$, while the dashed line shows the relation between $\alpha_3$ and $\alpha$ expected for small amplitude warps from the theory of \citetalias{ogilvie99} (Eq. \ref{eq:alpha3vsalpha}).}
\label{fig:alpha3}
\end{figure}

The resulting values of $\alpha_3$ as a function of $\alpha$ are plotted in Figure \ref{fig:alpha3} for Series 1 to 4 (that is, for the small amplitude cases with disc viscosity modelled through the artificial viscosity term). Given the large uncertainty at the lower resolution and lowest warp amplitude ($A=0.01$), error bars are shown only for the $A=0.05$ and higher resolution calculations for clarity (by way of comparison the error bars for the low resolution $A=0.01$ calculations are roughly a factor of two larger than for the corresponding low resolution $A=0.05$ results). The solid line shows the small amplitude and $\alpha \ll 1$ approximation $\alpha_{3} = 3/8$ (Eq. \ref{eq:alpha3const}), whilst the long-dashed line gives the expected relation for small warp amplitudes but to higher order in $\alpha$ (Eq.~\ref{eq:alpha3vsalpha}). Once again, provided the corrections for finite $\alpha$ are accounted for, we obtain a very good agreement between our numerical results and the theory, except for the lowest values of $\alpha$. The deviations seen at low $\alpha$ are most likely due to the effect of bulk viscosity in the code which affects the $l_{y}$ profile more strongly than either $l_{x}$ or $l_{z}$ (simply due to the low amplitude of $l_{y}$), and is more significant when the disc viscosity is low.

 We note briefly that --- by contrast with our earlier results for both $\alpha$ and $\alpha_{2}$ --- the calculations utilising the Navier-Stokes implementation of viscosity (Series 5 and 6) show a strong disagreement with both the modified artificial viscosity calculations and the theoretical curves -- the precession even reversing direction for $\alpha \lesssim 0.07$. The fits for series 6 --- i.e., the calculations that show good agreement in the $\alpha_{2}$ fits --- show the fit to $\alpha_{3}$ rise with $\alpha$ (from negative values) and then flatten to around $\alpha_{3} \approx 0.23$ at $\alpha \gtrsim 0.2$, in contrast to the results shown in Fig.~\ref{fig:alpha3}. The results for series 5 show a similar trend but with much lower fitted values. The errors to the fits are also significantly larger than for the artificial viscosity calculations. Whilst we can only speculate as to the reason for this disagreement, most likely it is an indication that higher resolution is needed (compared to using the modified artificial viscosity) to evaluate the nested first derivatives (Eqs. \ref{eq:dvterm} and \ref{eq:mom}) to the accuracy required in order to measure the precessional contribution. \citetalias{LP07} also found the precession in their simulations to depend strongly on details of the viscosity formulation.

\begin{figure}
\includegraphics[width=1.\columnwidth]{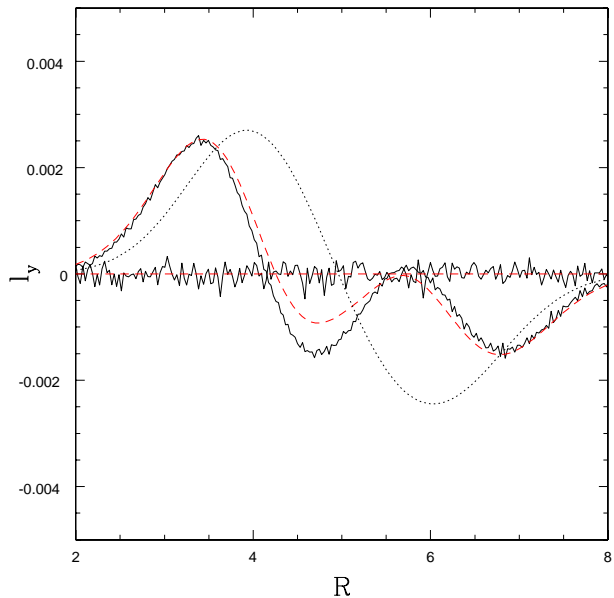}
\caption{Profile of $l_y$ at $t=500$ time units for the case $A=0.5$ and $\alpha=0.43$. The solid black lines refer to the SPH simulations, while the dashed red line show the result of the evolution of the diffusion plus precession code, with non linear warp parameter $\alpha_2$ and $\alpha_3$ computed based on \citetalias{ogilvie99} theory of warp propagation. For comparison, we also show with the dotted black line the profile at $t=500$ obtained from the simple 1D evolution model assuming a constant $\alpha_3= 0.22$, appropriate for this value of $\alpha$ in the linear regime ($\psi\ll 1$).}
\label{fig:prec_largeA}
\end{figure}

Finally, let us consider the internal precession for large amplitude warps ($A=0.5$, Series 7). In this case, as for the evaluation of the warp diffusion coefficient, we cannot simply associate a single value of $\alpha_3$ to our simulation, as it depends on the instantaneous value of $\psi$, which is a function of $R$ and $t$. However, we can still compare the profile of $l_y$ at a given time to the one expected from the non-linear theory of \citetalias{ogilvie99}. Once again, we stress that in this comparison we have not fitted any parameters, as the value of $\alpha$ is simply the input value in the simulation, while both $\alpha_2$ and $\alpha_3$ are a prescribed function (obtained from \citetalias{ogilvie99}) of $\alpha$, $\alpha_{\rm b}=5\alpha/3$ and $\psi$. The profile of $l_y$ for $\alpha=0.43$ and $A=0.5$ is shown in Fig. \ref{fig:prec_largeA} at $t=0$ and $t=500$ (in code units). The solid black lines show the results of the SPH simulations while the dashed red lines refer to the solution of the diffusion equation with added precession, where the coefficients are computed directly from the non linear theory of \citetalias{ogilvie99}. Note that, while in this large amplitude case the resulting shape of $l_y$ is a more complicated function than a simple oscillating function (as in the small amplitude case), the profile is reproduced surprisingly well by the \citetalias{ogilvie99} theory. To emphasize the importance of non-linear effects in this case, we also show with the dotted black line in Fig. \ref{fig:prec_largeA} the profile of $l_y$ at $t=500$ obtained from the 1D evolution code neglecting the effects of non-linearity and simply adopting a constant $\alpha_3=0.22$, that is the value of the precession coefficient for $\alpha=0.43$ in the small amplitude limit. One can thus clearly see that the non-linearity in the determination of $\alpha_3$  is essential in order to reproduce the correct precession of the disc. 

\section{Conclusions}
\label{sec:conclusions}

In this paper we have numerically tested the non-linear propagation of warps in thin and viscous accretion discs. To this end, we have run very high resolution SPH simulations of warped accretion discs, extending the previous work of \citetalias{LP07} to cover a much wider region of the parameter space. In some simulations, we have increased the numerical resolution with respect to \citetalias{LP07} by using ten times as many particles. We have also checked the effect of two different implementation of the disc viscosity. 

Our new and improved results correct upon the previous results of \citetalias{LP07}, who had found a disagreement between their simulations and the analytical theories of warp propagation. On the contrary, our results are in spectacular agreement with the non-linear theory of warp propagation of \citetalias{ogilvie99}. Some specific features of this theory, confirmed by our simulations, are worth recalling:

\begin{enumerate}

\item For moderate values of $\alpha\gtrsim 0.1$, the warp diffusion coefficient $\nu_2$ is not proportional to $1/\alpha$, where $\alpha$ is the disc viscosity coefficient, but follows the slightly more complex relation, Eq. (\ref{eq:prediction2}). The `standard' $1/\alpha$ behaviour is only recovered for smaller values of $\alpha$ \emph{and for small amplitude warps}. Note that a value of $\alpha\approx 0.1$ is expected based on observations of accreting binary systems \citep{king07}.

\item For large amplitude warps, the relation between $\nu_2$ and $\nu$ is much flatter than the $1/\alpha$ relation, corresponding to a more uniform (with respect to the disc viscosity) but also much less efficient diffusion of the warp at low $\alpha$ compared to the linear case. Our simulations, which are characterised by a warp amplitude $\psi\approx 1$ are reasonably well described by an almost constant $\alpha_2\approx 2.5$. 

\item In general, for non-linear warps, the warp diffusion coefficient is a function of the warp amplitude, which is itself a function of position and time. For a proper calculation of the warp evolution in a simple 1D diffusion code it is essential to include such dependence. We stress that this can and should be done in any warp diffusion code.

\item The non-linear theory also predicts the appearance of internal precessional torques. Also such torques are well described by the non-linear theory of \citetalias{ogilvie99} and can be easily included in 1D models by the simple addition of an extra term in the evolution equation, as discussed in the text.

\item For large warp amplitudes and small viscosity ($\psi>\sqrt{24}\alpha$) the evolution of the system is not well described by a simple diffusion equation and a full numerical approach is thus needed in such cases.

\end{enumerate}

All of the above aspects of warp propagation are reproduced faithfully in our simulations and can deeply affect the structure and evolution of the disc. We stress that their implementation (except for the last point) is relatively easy and recommended in simple 1D models of warped accretion disc evolution.

Before concluding, we should add a word of caution. All of the above results refer to warp propagation in a disc where viscosity is a standard Navier-Stokes viscosity, and in particular to the case where it is isotropic. Accretion disc viscosity is generally thought to be due to turbulence driven by some disc instability, such as the magneto-rotational instability \citep{balbus91} or gravitational instabilities \citep{LR04,lodatoNC}. In such cases, it is not obvious that the induced transport can be described in terms of an isotropic viscosity coefficient, and some of the above results, in particular concerning the precessional terms \citepalias{LP07}, might be affected.

\section*{Acknowledgements}
We thank Jim Pringle for several stimulating discussions and Gordon Ogilvie for providing us with the routine to compute the non-linear warp coefficients and for helpful comments on the paper. GL acknowledges the support of the Isaac Newton Institute for Mathematical Studies, where some of this work has been carried out. DJP acknowledges the support of a Monash fellowship. A substantial part of this work was also completed whilst DJP was funded by a UK Royal Society University Research Fellowship. Calculations were performed on zen, the SGI Altix ICE supercomputer at the University of Exeter and on the Monash Sun Grid. DJP thanks Dave Acreman at Exeter and Philip Chan and the e-research centre at Monash for ably managing the respective facilities. Figures~\ref{fig:warp3D}, \ref{fig:xsection} and \ref{fig:discbreak} were produced using \textsc{splash} \citep{splashpaper}, a visualisation tool for SPH data.

\bibliography{lodato,sph}

\begin{thebibliography}{}

\bibitem[\protect\citeauthoryear{Artymowicz \& Lubow}{Artymowicz \&
  Lubow}{1994}]{lubow94}
Artymowicz P.,  Lubow S.~H.,  1994, ApJ, 421, 651

\bibitem[\protect\citeauthoryear{Balbus \& Hawley}{Balbus \&
  Hawley}{1991}]{balbus91}
Balbus S.~A.,  Hawley J.~F.,  1991, ApJ, 376, 214

\bibitem[\protect\citeauthoryear{{Bardeen} \& {Petterson}}{{Bardeen} \&
  {Petterson}}{1975}]{bardeen75}
{Bardeen} J.~M.,  {Petterson} J.~A.,  1975, ApJ, 195, L65

\bibitem[\protect\citeauthoryear{Bate}{Bate}{1995}]{batephd}
Bate M.,  1995, PhD thesis, University of Cambridge, UK

\bibitem[\protect\citeauthoryear{{Bate}, {Lodato} \& {Pringle}}{{Bate}
  et~al.}{2010}]{BLP09}
{Bate} M.~R.,  {Lodato} G.,    {Pringle} J.~E.,  2010, MNRAS, 401, 1505

\bibitem[\protect\citeauthoryear{{Begelman}, {King} \& {Pringle}}{{Begelman}
  et~al.}{2006}]{begelman06b}
{Begelman} M.~C.,  {King} A.~R.,    {Pringle} J.~E.,  2006, MNRAS, 370, 399

\bibitem[\protect\citeauthoryear{{Benz}, {Cameron}, {Press} \& {Bowers}}{{Benz}
  et~al.}{1990}]{benzetal90}
{Benz} W.,  {Cameron} A.~G.~W.,  {Press} W.~H.,    {Bowers} R.~L.,  1990, ApJ,
  348, 647

\bibitem[\protect\citeauthoryear{{Brookshaw}}{{Brookshaw}}{1985}]{brookshaw85}
{Brookshaw} L.,  1985, Proceedings of the Astronomical Society of Australia, 6,
  207

\bibitem[\protect\citeauthoryear{{Chiang} \& {Murray-Clay}}{{Chiang} \&
  {Murray-Clay}}{2004}]{chiang04}
{Chiang} E.~I.,  {Murray-Clay} R.~A.,  2004, ApJ, 607, 913

\bibitem[\protect\citeauthoryear{{Cleary} \& {Monaghan}}{{Cleary} \&
  {Monaghan}}{1999}]{cm99}
{Cleary} P.~W.,  {Monaghan} J.~J.,  1999, J. Comp. Phys., 148, 227

\bibitem[\protect\citeauthoryear{{Dotti}, {Volonteri}, {Perego}, {Colpi},
  {Ruszkowski} \& {Haardt}}{{Dotti} et~al.}{2009}]{dotti09}
{Dotti} M.,  {Volonteri} M.,  {Perego} A.,  {Colpi} M.,  {Ruszkowski} M.,
  {Haardt} F.,  2009, MNRAS, p.~1795

\bibitem[\protect\citeauthoryear{{Espa{\~ n}ol} \& {Revenga}}{{Espa{\~ n}ol} \&
  {Revenga}}{2003}]{er03}
{Espa{\~ n}ol} P.,  {Revenga} M.,  2003, Phys. Rev. E, 67, 026705

\bibitem[\protect\citeauthoryear{{Flebbe}, {Muenzel}, {Herold}, {Riffert} \&
  {Ruder}}{{Flebbe} et~al.}{1994}]{flebbeetal94}
{Flebbe} O.,  {Muenzel} S.,  {Herold} H.,  {Riffert} H.,    {Ruder} H.,  1994,
  ApJ, 431, 754

\bibitem[\protect\citeauthoryear{{Herrnstein}, {Greenhill} \&
  {Moran}}{{Herrnstein} et~al.}{1996}]{herrnstein96}
{Herrnstein} J.~R.,  {Greenhill} L.~J.,    {Moran} J.~M.,  1996, ApJ, 468, L17

\bibitem[\protect\citeauthoryear{{King}, {Lubow}, {Ogilvie} \&
  {Pringle}}{{King} et~al.}{2005}]{KLOP}
{King} A.~R.,  {Lubow} S.~H.,  {Ogilvie} G.~I.,    {Pringle} J.~E.,  2005,
  MNRAS, 363, 49

\bibitem[\protect\citeauthoryear{{King} \& {Pringle}}{{King} \&
  {Pringle}}{2006}]{king06}
{King} A.~R.,  {Pringle} J.~E.,  2006, MNRAS, 373, L90

\bibitem[\protect\citeauthoryear{{King}, {Pringle} \& {Hofmann}}{{King}
  et~al.}{2008}]{KPH08}
{King} A.~R.,  {Pringle} J.~E.,    {Hofmann} J.~A.,  2008, MNRAS, 385, 1621

\bibitem[\protect\citeauthoryear{{King}, {Pringle} \& {Livio}}{{King}
  et~al.}{2007}]{king07}
{King} A.~R.,  {Pringle} J.~E.,    {Livio} M.,  2007, MNRAS, 376, 1740

\bibitem[\protect\citeauthoryear{{Larwood}, {Nelson}, {Papaloizou} \&
  {Terquem}}{{Larwood} et~al.}{1996}]{larwood96}
{Larwood} J.~D.,  {Nelson} R.~P.,  {Papaloizou} J.~C.~B.,    {Terquem} C.,
  1996, MNRAS, 282, 597

\bibitem[\protect\citeauthoryear{{Lodato}}{{Lodato}}{2007}]{lodatoNC}
{Lodato} G.,  2007, Nuovo Cimento Rivista Serie, 30, 293

\bibitem[\protect\citeauthoryear{{Lodato} \& {Pringle}}{{Lodato} \&
  {Pringle}}{2006}]{LP06}
{Lodato} G.,  {Pringle} J.~E.,  2006, MNRAS, 368, 1196

\bibitem[\protect\citeauthoryear{{Lodato} \& {Pringle}}{{Lodato} \&
  {Pringle}}{2007}]{LP07}
{Lodato} G.,  {Pringle} J.~E.,  2007, MNRAS, 381, 1287

\bibitem[\protect\citeauthoryear{Lodato \& Rice}{Lodato \& Rice}{2004}]{LR04}
Lodato G.,  Rice W. K.~M.,  2004, MNRAS, 351, 630

\bibitem[\protect\citeauthoryear{{Lubow} \& {Ogilvie}}{{Lubow} \&
  {Ogilvie}}{2000}]{lubow00}
{Lubow} S.~H.,  {Ogilvie} G.~I.,  2000, ApJ, 538, 326

\bibitem[\protect\citeauthoryear{{Lubow}, {Ogilvie} \& {Pringle}}{{Lubow}
  et~al.}{2002}]{lubow02}
{Lubow} S.~H.,  {Ogilvie} G.~I.,    {Pringle} J.~E.,  2002, MNRAS, 337, 706

\bibitem[\protect\citeauthoryear{{Martin}, {Pringle} \& {Tout}}{{Martin}
  et~al.}{2007}]{martin07b}
{Martin} R.~G.,  {Pringle} J.~E.,    {Tout} C.~A.,  2007, MNRAS, 381, 1617

\bibitem[\protect\citeauthoryear{{Martin}, {Pringle} \& {Tout}}{{Martin}
  et~al.}{2009}]{martin09}
{Martin} R.~G.,  {Pringle} J.~E.,    {Tout} C.~A.,  2009, MNRAS, 400, 383

\bibitem[\protect\citeauthoryear{{Martin}, {Reis} \& {Pringle}}{{Martin}
  et~al.}{2008}]{martin08b}
{Martin} R.~G.,  {Reis} R.~C.,    {Pringle} J.~E.,  2008, MNRAS, 391, L15

\bibitem[\protect\citeauthoryear{{Martin}, {Tout} \& {Pringle}}{{Martin}
  et~al.}{2008}]{martin08a}
{Martin} R.~G.,  {Tout} C.~A.,    {Pringle} J.~E.,  2008, MNRAS, 387, 188

\bibitem[\protect\citeauthoryear{{Monaghan}}{{Monaghan}}{1989}]{monaghan89}
{Monaghan} J.~J.,  1989, J. Comp. Phys., 82, 1

\bibitem[\protect\citeauthoryear{Monaghan}{Monaghan}{1992}]{monaghan92}
Monaghan J.~J.,  1992, ARA\&A, 30, 543

\bibitem[\protect\citeauthoryear{{Monaghan}}{{Monaghan}}{1997}]{monaghan97}
{Monaghan} J.~J.,  1997, J. Comp. Phys., 136, 298

\bibitem[\protect\citeauthoryear{{Monaghan}}{{Monaghan}}{2005}]{monaghan05}
{Monaghan} J.~J.,  2005, Rep. Prog. Phys., 68, 1703

\bibitem[\protect\citeauthoryear{{Morris} \& {Monaghan}}{{Morris} \&
  {Monaghan}}{1997}]{mm97}
{Morris} J.~P.,  {Monaghan} J.~J.,  1997, J. Comp. Phys., 136, 41

\bibitem[\protect\citeauthoryear{Murray}{Murray}{1996}]{murray96}
Murray J.~R.,  1996, MNRAS, 279, 402

\bibitem[\protect\citeauthoryear{{Nelson} \& {Papaloizou}}{{Nelson} \&
  {Papaloizou}}{1999}]{nelson99}
{Nelson} R.~P.,  {Papaloizou} J.~C.~B.,  1999, MNRAS, 309, 929

\bibitem[\protect\citeauthoryear{{Nelson} \& {Papaloizou}}{{Nelson} \&
  {Papaloizou}}{2000}]{nelson00}
{Nelson} R.~P.,  {Papaloizou} J.~C.~B.,  2000, MNRAS, 315, 570

\bibitem[\protect\citeauthoryear{{Ogilvie}}{{Ogilvie}}{1999}]{ogilvie99}
{Ogilvie} G.~I.,  1999, MNRAS, 304, 557

\bibitem[\protect\citeauthoryear{{Ogilvie}}{{Ogilvie}}{2000}]{ogilvie00}
{Ogilvie} G.~I.,  2000, MNRAS, 317, 607

\bibitem[\protect\citeauthoryear{{Ogilvie}}{{Ogilvie}}{2006}]{ogilvie06}
{Ogilvie} G.~I.,  2006, MNRAS, 365, 977

\bibitem[\protect\citeauthoryear{{Ogilvie} \& {Dubus}}{{Ogilvie} \&
  {Dubus}}{2001}]{ogilviedubus}
{Ogilvie} G.~I.,  {Dubus} G.,  2001, MNRAS, 320, 485

\bibitem[\protect\citeauthoryear{{Papaloizou} \& {Lin}}{{Papaloizou} \&
  {Lin}}{1995}]{paplin95}
{Papaloizou} J.~C.~B.,  {Lin} D.~N.~C.,  1995, ApJ, 438, 841

\bibitem[\protect\citeauthoryear{{Papaloizou} \& {Pringle}}{{Papaloizou} \&
  {Pringle}}{1983}]{pappringle83}
{Papaloizou} J.~C.~B.,  {Pringle} J.~E.,  1983, MNRAS, 202, 1181

\bibitem[\protect\citeauthoryear{{Papaloizou}, {Terquem} \& {Lin}}{{Papaloizou}
  et~al.}{1998}]{papaloizou98}
{Papaloizou} J.~C.~B.,  {Terquem} C.,    {Lin} D.~N.~C.,  1998, ApJ, 497, 212

\bibitem[\protect\citeauthoryear{{Perego}, {Dotti}, {Colpi} \&
  {Volonteri}}{{Perego} et~al.}{2009}]{perego09}
{Perego} A.,  {Dotti} M.,  {Colpi} M.,    {Volonteri} M.,  2009, MNRAS, 399,
  2249

\bibitem[\protect\citeauthoryear{{Price}}{{Price}}{2004}]{price04}
{Price} D.~J.,  2004, PhD thesis, University of Cambridge, Cambridge, UK.
  astro-ph/0507472

\bibitem[\protect\citeauthoryear{{Price}}{{Price}}{2007}]{splashpaper}
{Price} D.~J.,  2007, Publ. Astron. Soc. Aust., 24, 159

\bibitem[\protect\citeauthoryear{{Price} \& {Federrath}}{{Price} \&
  {Federrath}}{2010}]{pf10}
{Price} D.~J.,  {Federrath} C.,  2010, MNRAS, submitted

\bibitem[\protect\citeauthoryear{{Price} \& {Monaghan}}{{Price} \&
  {Monaghan}}{2004}]{pm04b}
{Price} D.~J.,  {Monaghan} J.~J.,  2004, MNRAS, 348, 139

\bibitem[\protect\citeauthoryear{{Price} \& {Monaghan}}{{Price} \&
  {Monaghan}}{2007}]{pm07}
{Price} D.~J.,  {Monaghan} J.~J.,  2007, MNRAS, 374, 1347

\bibitem[\protect\citeauthoryear{{Pringle}}{{Pringle}}{1992}]{pringle92}
{Pringle} J.~E.,  1992, MNRAS, 258, 811

\bibitem[\protect\citeauthoryear{{Pringle}}{{Pringle}}{1996}]{pringle96}
{Pringle} J.~E.,  1996, MNRAS, 281, 357

\bibitem[\protect\citeauthoryear{{Scheuer} \& {Feiler}}{{Scheuer} \&
  {Feiler}}{1996}]{scheuer96}
{Scheuer} P.~A.~G.,  {Feiler} R.,  1996, MNRAS, 282, 291

\bibitem[\protect\citeauthoryear{Shakura \& Sunyaev}{Shakura \&
  Sunyaev}{1973}]{shakura73}
Shakura N.~I.,  Sunyaev R.~A.,  1973, A\&A, 24, 337

\bibitem[\protect\citeauthoryear{{Wijers} \& {Pringle}}{{Wijers} \&
  {Pringle}}{1999}]{wijers99}
{Wijers} R.~A.~M.~J.,  {Pringle} J.~E.,  1999, MNRAS, 308, 207

\end{thebibliography}

\label{lastpage}
\end{document}